\documentstyle[emulateapj]{article}

\def\hii{\protect{\sc H\thinspace ii}}

\def\oiii{[\ion{O}{3}]}
\def\oii{[\ion{O}{2}]}
\def\siii{[\ion{S}{3}]}
\def\sii{[\ion{S}{2}]}
\def\siv{[\ion{S}{4}]}
\def\sivlam{[\ion{S}{4}]$\lambda10.5\mu$}
\def\neiii{[\ion{Ne}{3}]}

\def\op{O$^+$}
\def\opp{O$^{++}$}
\def\spp{S$^{++}$}
\def\sp3{S$^{+3}$}

\def\etal{et\thinspace al.~}
\def\Ha{H$\alpha$}
\def\Hb{H$\beta$}

\def\ha{\rm H\alpha}
\def\hb{\rm H\beta}

\def\Zsol{{\rm\,Z_\odot}}
\def\te{$T_e$}
\def\tlo{$T$(\op)}
\def\thi{$T$(\opp)}
\def\tstar{$T_\star$}

\def\lam{$\lambda$}
\def\cc{{\rm\,cm^{-3}}}
\def\ergs{{\rm\,erg\ s^{-1}}}
\def\logOH{log(O/H)}

\received{17 January 2000}
\accepted{17 March 2000}

\slugcomment{Submitted to ApJ\  17 January 2000; accepted 17 March 2000}

\lefthead{Oey and Shields}
\righthead{Calibration of Abundance Diagnostics}

\begin{document}

\title{Calibration of Nebular Emission-Line Diagnostics: 
	II.  Abundances}

\author{M. S. Oey}
\affil{Space Telescope Science Institute, 3700 San Martin Drive,
	Baltimore, MD   21218, USA; oey@stsci.edu}
\author{J. C. Shields}
\affil{Ohio University, Dept. of Physics and Astronomy, Clippinger
	Research Labs. 251B, Athens, OH   45701-2979, USA; 
	shields@helios.phy.ohiou.edu}

\begin{abstract}
We examine standard methods of measuring nebular chemical abundances,
including estimates based on direct \te\ measurements, and also
emission-line diagnostics.  We use observations of the LMC \hii\ regions
DEM L199, DEM L243, DEM L301, and DEM L323, whose ionizing stars have
classifications ranging from O7 to WN3.
Following common practice, we assume a two-zone \te\
structure given by \thi\ and \tlo\ to compute ionic abundances.
We compare with photoionization models tailored to the observed properties 
of the individual objects, and emphasize
the importance of correctly relating \te\ in the two zones, which can
otherwise cause errors of $\sim$0.2 dex in abundance estimates.  The
data show no spatial variations or local metallicity enhancements to
within 0.1 -- 0.15 dex in any of the objects, notably including DEM
L199, which hosts three WR stars.

Our data agree well with both the modeled $R23$ and $S23$ abundance
diagnostics for O and S.  We present the first theoretical tracks for
$S23$, which are in excellent agreement with a larger available
dataset.  However, contrary to earlier suggestions, $S23$ is 
much {\it more} sensitive to the ionization parameter ($U$) than is $R23$.
This occurs because $S23$ does not sample \ion{S}{4}, which is often
a significant population.  We therefore introduce
$S234\equiv$(\sii$+$\siii$+$\siv)/\Hb, 
and demonstrate that it is virtually independent of 
$U$.  Predicted and observed spatial variations in $S234$ are 
thus dramatically decreased in contrast to $S23$.  The intensity of
\siv\lam10.5$\mu$ can be easily estimated from the simple
correspondence between \siv/\siii\ and \oiii/\oii.  Using this method to 
estimate $S234$ for data in the literature yields excellent agreement
with our model tracks, hence we give a theoretical calibration for
$S234$.  Our models show that the double-valued structure of $S23$ and
$S234$ remains an important problem as for $R23$, and presently {\it we consider
calibrations of these S diagnostics reliable only at $Z\lesssim
0.5\Zsol$}.  However, the slightly larger dynamic range and excellent
compatibility with theoretical predictions suggest the S
parameters to be more effective abundance diagnostics than $R23$.
\end{abstract}


\section{Introduction}

The emission-line signatures of \hii\ regions are a powerful and
widely-used indicator of galactic abundances.  This is especially true in
distant galaxies where most stellar abundance probes cannot be
employed.  Furthermore, the spectral properties of \hii\ regions also give
important diagnostics of the ionizing stellar population, such as
effective temperature and numbers of stars.  Given the wide use of
such nebular diagnostics, it is vital to test and calibrate them using
\hii\ regions with known characteristics and ionizing stellar
populations.

Along with a companion paper (Oey {\etal}2000; hereafter Paper~I), we
report here on a detailed study for this purpose, of four \hii\
regions in the Large Magellanic Cloud (LMC).  The OB associations in
all four of these \hii\ regions have been examined in detail and classified,
thereby strongly constraining the ionizing energy distributions.
In addition, the LMC's proximity and orientation with respect to the
Galaxy also permit a detailed understanding of the nebular morphology.
Thirdly, the abundances can be accurately determined.  We therefore
have high-quality information on the three principal parameters that
determine the nebular emission:  stellar effective temperature
(\tstar), the ionization parameter ($U$) that relates ionizing flux to 
gas density, and metallicity ($Z$).

Table~\ref{sample} gives a brief
summary of our objects, which can be examined in greater detail in
Paper~I.  The first two columns in Table~\ref{sample} give the
\hii\ region identification in the Davies, Elliott, \& Meaburn (1976)
and Henize (1956) \Ha\ catalogs, respectively; the third column
identifies the parent OB association from Lucke \& Hodge (1970).
Column four shows the spectral type of the dominant ionizing stars, 
as classified by the references in column 8.
Column five lists the nebular \Ha\ luminosity from Oey \& Kennicutt
(1997), and column 6 shows the adopted value from Paper~I, of the
inner radius of the gas distribution, as a fraction of the Str\"omgren
radius $R_{\rm S}$.  This gives some indication of the nebular morphology.
Finally, column 7 indicates the presence of significant shock
excitation:  DEM L243 includes an embedded or superimposed supernova
remnant (SNR), and DEM L301 also shows evidence of significant shock
activity (Paper I).  While shocks may be present in the other objects
as well, their contribution to the nebular emission is unimportant. 

The detailed presentation of the objects is given in
Paper~I, including narrow-band images in \Ha, \oiii, and \sii, and
spectroscopic observations over the
wavelength range 3500 -- 9200 \AA.  For each object, we observed two to
three stationary, spatially-resolved slit positions.  For all of the
objects except DEM L301, we also obtained spatially integrated
observations by scanning the long slit across the nebulae.  The scanned
data should therefore resemble typical observations of such \hii\ regions
at distances of 10 -- 20 Mpc.  As seen in Table~\ref{sample}, the
sample spans a range in dominant stellar spectral types, from O7 to
early Wolf-Rayet (WR).  There is also variety in the morphology of the
objects, ranging from near perfect Str\"omgren sphere (DEM L323), to
extreme shell structure (DEM L301), to highly complex (DEM L199).

Paper~I provides a detailed analysis of the spatially resolved
emission-line ratios with respect to the sequence in \tstar\ 
represented by these objects, and a comparison with photoionization
models using the current generation of stellar atmosphere models for
both O stars and early WR stars.  In general we found a gratifyingly
high level of agreement, largely supporting the CoStar energy
distributions of Schaerer \& de Koter (1997) and early WR atmospheres
of Schmutz, Leitherer, \& Gruenwald (1992) and Hamann \& Koesterke (1998).
In conjunction with data from Kennicutt {\etal}(2000), we provide
a first, empirical calibration of nebular diagnostics for the dominant
\tstar.  In addition to the well-known $\eta^\prime$ radiation
softness parameter of V\'\i lchez \& Pagel (1988), we introduce 
[\ion{Ne}{3}]\lam3869/\Hb\ as a similar diagnostic, which is more
robust to nebular conditions and is sensitive for a higher range of \tstar.

We also presented in Paper~I the spatially-resolved behavior of the
O abundance parameter (Pagel {\etal}1979),
\begin{equation}\label{eqR23}
R23 \equiv \rm\frac{[O\thinspace II]\lambda3727 + 
	[O\thinspace III]\lambda\lambda4959,5007}{\hb} \quad ;
\end{equation}
and S abundance parameter (V\'\i lchez \& Esteban 1996; Christensen
{\etal}1997; D\'\i az {\etal}1999), 
\begin{equation}\label{eqS23}
S23 \equiv \rm\frac{[S\thinspace II]\lambda6724 + 
	[S\thinspace III]\lambda\lambda9069,9532}{\hb} \quad ,
\end{equation}  
where we designate the \sii\ lines $\lambda6716 + \lambda6732$ as
\lam6724, analogous to \oii\lam3727.  We confirmed that both
observationally and theoretically, $R23$ remains spatially uniform
across the nebulae.  In contrast, for uniform abundances, models
of $S23$ vary across the nebulae, showing lower values in central
regions.  The observations clearly reflect this pattern, which
is caused by the missing ionization stage of \ion{S}{4}.  We address
this issue in detail below in this paper.  As is well-known, $R23$ and $S23$
are also sensitive to \tstar, and these effects are also shown in
Paper~I for our objects.

In this paper, we present the abundance determinations for our sample,
for both the spatially-resolved and scanned longslit observations.
We examine conventional assumptions for the nebular electron
temperature (\te) structure and 
resulting ionic and total abundance determinations.
We then explore the metallicity diagnostics $R23$ and $S23$ in more
detail, and introduce another diagnostic, $S234$.  As 
before, we use the photoionization code {\sc Mappings~II} (Sutherland
\& Dopita 1993) in conjunction with CoStar stellar atmosphere models
(Schaerer \& de Koter 1997) for O stars.

\bigskip\bigskip
\section{Direct abundance determinations}

We derived abundances from the measured line fluxes in Paper~I using
standard techniques.  For calculation of line emissivities for
elements heavier than helium, we used the five-level atom code {\sc Five\_L}
as implemented in {\sc Stsdas} version 1.3.5 (Shaw \& Dufour 1995).  We
initially obtained an estimate of electron density from the
[\ion{S}{2}] $\lambda$6716/$\lambda$6731 line intensity ratio,
assuming a default $T_e=10^4$~K.  The densities 
are almost all $\lesssim 100\ \cc$.

\subsection{Temperature structure}

The assumed \te\ structure, and thereby,
ionization structure of the 
\hii\ region, can generate substantial uncertainties in 
even ``direct'' abundance determinations from typical optical
emission-line spectra.  The \te\ structure is in turn a
function of density structure and stellar atmosphere models.  
Peimbert's (1967) $t^2$ formulation for temperature fluctuations is
probably the best-known example of addressing this problem, which can
be used for both small-scale and large-scale variations in \te.

We consider here the large-scale temperature structure.  As discussed
below, our photoionization models tend to overestimate \te,
suggesting that small-scale fluctuations are not significant.
Standard abundance determinations adopt either a single characteristic
\te\ or a two-zone model for \te, and these assumptions can also 
cause substantial errors in abundance determinations.  Garnett (1992)
provides one of the more thorough investigations of this issue, which is a 
well-known problem for elements like S that do not conform well to a
two-zone model.  However, if the adopted zone temperatures
do not adequately characterize the respective regions, then the
inferred abundance can be significantly in error even for elements
like O that are well-described by a two-zone model.  This can be a
problem especially in cooler nebulae with $T_e \lesssim 10,000$ K,
that have strong temperature gradients (Garnett 1992; Stasi\'nska 1980).

We assumed here a two-zone temperature structure for our nebulae, such
that a common \te\ was adopted for excitation of \ion{O}{3}, \ion{Ne}{3},
\ion{S}{3}, and \ion{Ar}{3}; while a separate \te\ was adopted
for the excitation of \ion{N}{2}, \ion{O}{2}, and
\ion{S}{2}.  We used the same electron density for both zones.

For the high ionization zone, we adopted $T$(\opp), the
volume-averaged \te\ for the \opp\ population.  This is obtained
essentially directly from the observed $T$\oiii, the temperature
inferred from the [\ion{O}{3}] $\lambda$4363/($\lambda$4959 + 
$\lambda$5007) line intensity ratio.  A temperature can also be
obtained in principle from the [\ion{S}{3}]
$\lambda$6312/($\lambda$9069 + \lam9532) ratio.  However, the
auroral line flux in this case was often described by a large
fractional uncertainty, which introduced correspondingly large random
errors in the resulting abundances; there is also some possibility
that the near-IR lines may be affected by telluric absorption
(e.g., Stevenson 1994).  We consequently chose to adopt \thi$=T$\oiii\
for the high-ionization zone, including the S$^{++}$
region.  Garnett (1992) has suggested on the basis of nebular models
that the relation between ion-weighted temperatures for S$^{++}$ and
O$^{++}$ is linear, but with a slope differing from unity.  For
objects in the current study, 
$T$[\ion{O}{3}] values are generally close to 10,000~K, at which
point Garnett's work suggests that the ion-weighted $T$(\spp) and \thi\
should be similar; thermal effects of dust in the nebula may also be
expected to reduce the contrast between the two temperatures (Shields
\& Kennicutt 1995), lending support to use of a common value.

For the low-ionization zone, \te\ can be
obtained directly from the [\ion{S}{2}]($\lambda$4069 +
$\lambda$4076)/($\lambda$6716 + $\lambda$6731) ratio, but we found
that this option frequently suffered from a low signal-to-noise ratio.
It is often standard practice to derive the lower-ionization
\tlo\ from the higher-ionization \thi\ using an analytic
relation (e.g., Campbell, Terlevich, \& Melnick 1986; Pagel
{\etal}1992), whose accuracy clearly affects that of the 
abundance determination.  We investigated several alternatives before
adopting a prescription for \tlo.  In deciding what relation to adopt, our 
principal criterion was that the abundances input to {\sc Mappings}
photoionization models for the individual objects (Paper~I) should be
consistent with those inferred from the output emission-line spectra.
For example, when using the expression employed by Pagel {\etal}(1992,
their equation 6), our resultant \hii\ region model spectra implied an O
abundance that was $\sim 0.2$ dex lower than the assumed input, a
substantial departure from self-consistency!  

\begin{figure*}
\epsscale{1.7}
\plotone{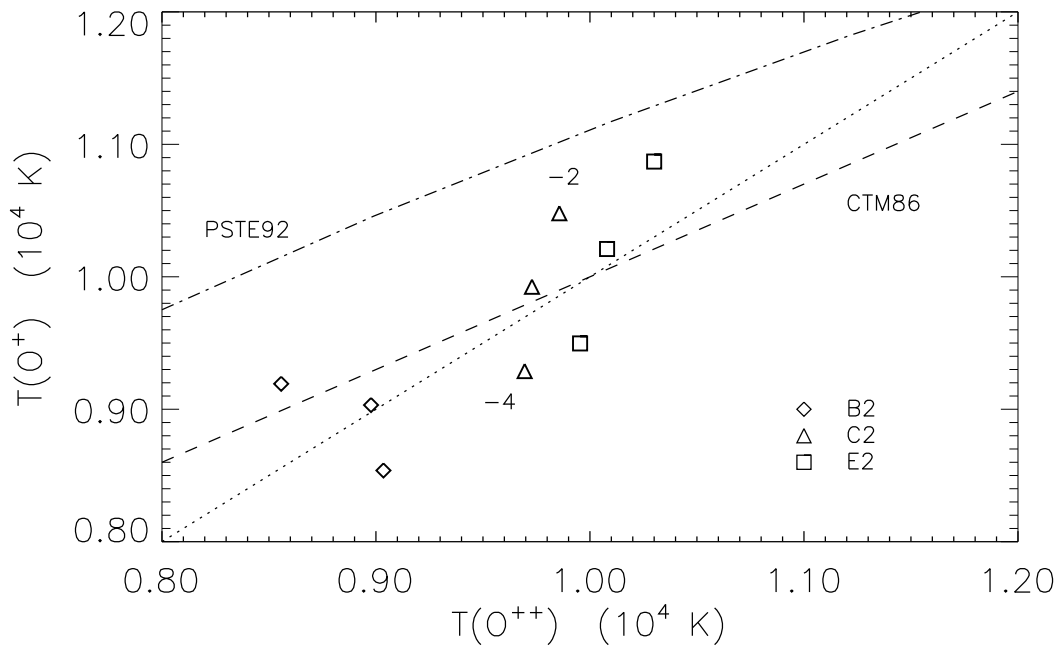}
\vspace*{-3.7in}
\caption{\tlo\ vs. \thi\ from photoionization models.  The symbols
correspond to models using the indicated CoStar stellar atmospheres
(see text).  The locus of results for $\log U=-2,\ -3,$ and $-4$ fall
in the sense indicated for the C2 runs.
We overplot dashed and dot-dashed lines that show the relations of
Campbell {\etal}(1986) and Pagel {\etal}(1992), respectively; the
dotted line shows \tlo$=$\thi.
\label{TO3O2}}
\end{figure*}

The relationship between
\thi\ and \tlo\ is model-dependent, and varies with ionization
parameter and metallicity.  Figure~\ref{TO3O2} shows \tlo\ vs. \thi\ for {\sc
Mappings} photoionization models at $Z=0.3\Zsol$, with an inner radius
of 0.4$R_{\rm S}$.  The different symbols correspond to different
CoStar stellar atmospheres as indicated,
with B2, C2, and E2 corresponding to spectral types O8--O9, O6--O7,
and O3--O4, respectively.  We show models with $\log U=-2,\ -3,$ and
$-4$.  The dashed and dot-dashed lines show the relation between \tlo\ and
\thi\ from Campbell {\etal}(1986; see also Garnett 1992), and Pagel
{\etal}(1992), respectively, while the dotted line delineates \tlo$=$\thi.

The Pagel {\etal}relation deviates significantly from the models at these
temperatures, while the Campbell {\etal}relation shows reasonable
agreement in slope for \thi\ $> 10,000$ K, falling close to the
theoretical predictions for $\log U = -3$.  At lower temperatures, the
models are more consistent with an isothermal nebula.
We consequently adopted the formulation,
\begin{equation}\label{garneq}
T({\rm O^+}) = 
\left\{
\begin{array}{ll}
	0.7\ T({\rm O^{++}}) + 3000\ {\rm K}\ ,
	& T({\rm O^{++}}) > 10,000\ {\rm K}  \\

	T({\rm O^{++}})\ , 
	& T({\rm O^{++}}) < 10,000\ {\rm K} 
\end{array}
\right.
\end{equation}
where $T$(\opp)$=T$[\ion{O}{3}] as described above.  The relation for
$T$(\opp) $> 10,000$ K is that given by Campbell {\etal}\
Equation~\ref{garneq} yields consistent input and
output abundances to $\lesssim 0.05$ dex, except for Ne, whose modeled
line ratios are persistently discrepant with the observations
(Paper~I; see below).
We caution that Figure~\ref{TO3O2} and
equation~\ref{garneq} are optimized in the parameter space relevant to
our LMC targets.  Different formulations may be more appropriate at
other metallicities and ionization parameters; it is beyond the scope
of this work to fully examine this issue.  However, it is clear that
some care is necessary in choosing a relation between \thi\ and \tlo.

While we ensured consistency between the input and output abundances
to the photoionization models, one worrisome problem that remains
unresolved is the overprediction of \te\ in comparison to the
observations, as presented in Paper~I.  To briefly summarize, the
predicted $T$\oiii\ is greater than that inferred from the observed
\oiii\lam4363/\lam5007 by 850 K in DEM L323 and 1500 K in DEM L199.
The problem may also exist in the other two objects, DEM L243 and DEM
L301, but the fainter emission from \lam4363 in these lower-excitation nebulae
prevented any useful constraints.  We also found that the discrepancy
persisted when using Hummer \& Mihalas (1970) stellar atmosphere
models and the {\sc Cloudy} (Ferland 1998) photoionization code.
Since \thi\ and $T$\oiii\ agree to $< 1$\% in the models, the effect
is not caused by non-collisional excitation of \lam4363 in the models.
As discussed in Paper~I, the temperature discrepancy occurs in lines
of sight across the entire nebulae, and is in the {\it opposite} sense
of that expected from small-scale temperature fluctuations (Peimbert
1967).  Our sample therefore shows {\it no} evidence that such \te\
fluctuations systematically bias our abundance determinations.  Mathis
(1995) discusses evidence for and against the general existence of significant
\te\ fluctuations.

\subsection{Ionic abundances}

Determining ionic abundances relative to H requires calculation
of the H$\beta$ emissivity, which is also temperature dependent, but
originates in both the high- and low-ionization zones.  We adopted an
intermediate temperature for calculation of the H$\beta$ emissivity,
representing an average of \thi\ and \tlo, weighted by the relative
abundances of \opp\ and \op.  The same intermediate
temperature was used for computing 
He emissivities, using expressions taken from Benjamin et
al. (1999).  This construction was used to obtain ionic abundances
from the line measurements in Paper I, which were corrected for
reddening and, where necessary, underlying Balmer absorption.  The
abundance results for different nebular lines of a single ion were combined,
weighted by the variance of the line fluxes.

Total abundances were obtained from the ionic abundances following
the standard practice of using ionization correction factors (ICFs)
to allow for unobserved ionization stages.  We 
employ the following relations:
\begin{equation}\label{oxygen_eq}
\rm \frac{O}{H} = \frac{O^+ + O^{++}}{H^+} \quad ,
\end{equation}
where the lack of detectable \ion{He}{2}\lam4686 emission indicates no
significant O$^{+3}$ population;
\begin{equation}
\rm \frac{N}{H} = \frac{N^+}{O^+}\cdot\frac{O}{H} 
\end{equation}
(Peimbert \& Costero 1969; Garnett 1990);
and
\begin{equation}
\rm \frac{S}{H} = \biggl[\frac{S^+ + S^{++}}{H^+}\biggr] \Bigl / ICF(S^{+3}) \quad ,
\end{equation}
where
\begin{equation}
\rm ICF(S^{+3}) = \biggl[ 1 - 
	\biggl( 1 - \frac{O^+}{O}\biggr)^\alpha \biggr]^{1/\alpha} \quad ,
\end{equation}
with $\alpha = 2.5$, corrects for unobserved S$^{+3}$ ions (Garnett
1989; Stasi\'nska 1978).  For the noble gases, we adopt 
\begin{equation}\label{eqNe}
\rm \frac{Ne}{H} = \frac{Ne^{++}}{O^{++}}\cdot\frac{O}{H}
\end{equation}
(Peimbert \& Costero 1969; Simpson {\etal}1995), and
\begin{equation}\label{argon_eq}
\rm \frac{Ar}{H} = \frac{Ar^{++}}{S^+ + S^{++}}\cdot\frac{S}{H}
\end{equation}
(Garnett {\etal}1997).  Most of our objects are ionized by
early-type stars (Table~\ref{sample}), thereby fully ionizing He, but
not exhibiting detectable \ion{He}{2}.  Thus our default relation for
He is simply,
\begin{equation}\label{helium_eq}
\rm \frac{He}{H} = \frac{He^+}{H^+} \quad .
\end{equation}
%
For DEM L243, which is ionized by O7 stars, the
\ion{He}{1} \lam5876/\Hb\ line ratios suggest 
that He is not fully ionized throughout the nebula (Paper~I;
see below).  For this object, we experimented with 
the expression for He/H given by Peimbert \& Torres-Peimbert (1977; their
equation~15), that includes an ICF for He$^0$; however, this
prescription yields He 
abundances that are at least 0.1 dex higher than the rest of the
objects in the sample.  We therefore give results for DEM L243 using
equation~\ref{helium_eq} and note that these should be treated as
lower limits to the true abundance.  It is worth remarking that our
investigation supports the results of Baldwin {\etal}(1991), who
find that the Orion He abundance was slightly overestimated by
Peimbert \& Torres-Peimbert as a result of their ICF for He$^0$.

The theoretical support for use of equations \ref{oxygen_eq} --
\ref{argon_eq} derives mostly from models that consider the integrated
properties of \hii\ regions (e.g., Garnett 1992; Mathis 1985;
Stasi\'nska 1978).  Thus, observations that sample only a
small ``pencil-beam'' through a nebula may not yield 
the correct abundances if such an analysis scheme is used;
Gruenwald \& Viegas (1992) have discussed this problem in detail.
However, previous observational studies have found generally good
agreement in elemental abundances derived at variable positions across
individual nebulae, using the integrated-spectrum methods (e.g.,
D\'\i az {\etal}1987; Masegosa, Moles, \& del Olmo
1991; Gonz\'alez-Delgado {\etal}1994).  Such agreement was found even
when substantial excitation gradients were seen within the \hii\
region.  In the present study, we can
directly compare abundances obtained from the integrated spectrum of a
nebula with those calculated by the same means from small-aperture
measurements at a variety of radii.  We show below in \S 2.3.1 that we
find a high degree of consistency between the different aperture
measurements for the same source, including the integrated spectrum.
These results support the validity of using
integrated-spectrum prescriptions when only a part of the \hii\ region
is observed.  
Apparently the characteristic physical parameters determined for the
two ionization zones are adequate to determine the ionic abundances to
high accuracy in our objects.



\subsection{Abundance results}

We used the prescriptions above to compute abundances for
both the spatially resolved and scanned, spatially integrated observations
for each object.  Table~\ref{stadat} presents results for the
individual, spatially resolved observations, along with error estimates.
These were obtained from the uncertainties in measured fluxes,
using a Monte Carlo method similar to that described by Kobulnicky \&
Skillman (1996).  Input flux values were modified randomly by addition
of Gaussian noise with amplitude corresponding to 1-$\sigma$
measurement uncertainties, for a total of 5000 iterations per
spectrum.  A few points with low signal-to-noise ratios yielded unphysical
solutions in the \te\ and abundance determinations, and these were
discarded from the sample.  The standard deviation of the resulting abundance
distribution was adopted as the final uncertainty.

Table~\ref{meandat} presents the mean values of the spatially-resolved
measurements, weighted inversely by the variances.  
We also list the corresponding formal uncertainty in the mean.
Note that the weighting scheme and quoted errors assume
that actual variations are negligible in the quantities listed in
Table~\ref{meandat}, and that 
the scatter results from a normal distribution of measurement errors.  
As described below, we do not find significant evidence for
abundance fluctuations; however, it is possible that the measurement
errors may not be a strictly normal distribution.  
The listed errors in Table~\ref{meandat} are therefore likely
to somewhat underestimate the true uncertainties, nor do these include
systematic errors.  

Regarding the abundances of Ne, it is important to note that there is
probably a substantial systematic error in the values obtained in
Table~\ref{meandat}, and also later in Table~\ref{scandat}.  As
discussed in Paper~I, the observed line intensities for Ne are
systematically discrepant with the tailored models.  Although earlier
generation stellar atmospheres caused \neiii\ intensities to be
underpredicted, we now find a modest overprediction.  The discrepancy
can be resolved by reducing log(Ne/H) by 0.2 -- 0.3 dex.  However,
since we do not understand the cause of the discrepancy, we have
chosen to list the Ne abundance derived from the standard relations.
It appears likely that errors in the stellar atmospheres are
responsible for much of the problem (Paper~I).  Peimbert (1993) also
emphasizes the uncertainty in 
equation~\ref{eqNe}.  The high ionization potential (40.96 eV)
required for \ion{Ne}{3} gives it outstanding potential for probing
hot stellar ionizing sources, so it is highly desireable to
resolve the uncertainties regarding its emission.

\subsubsection{Spatial uniformity}

\begin{figure*}
\epsscale{1.7}
\plotone{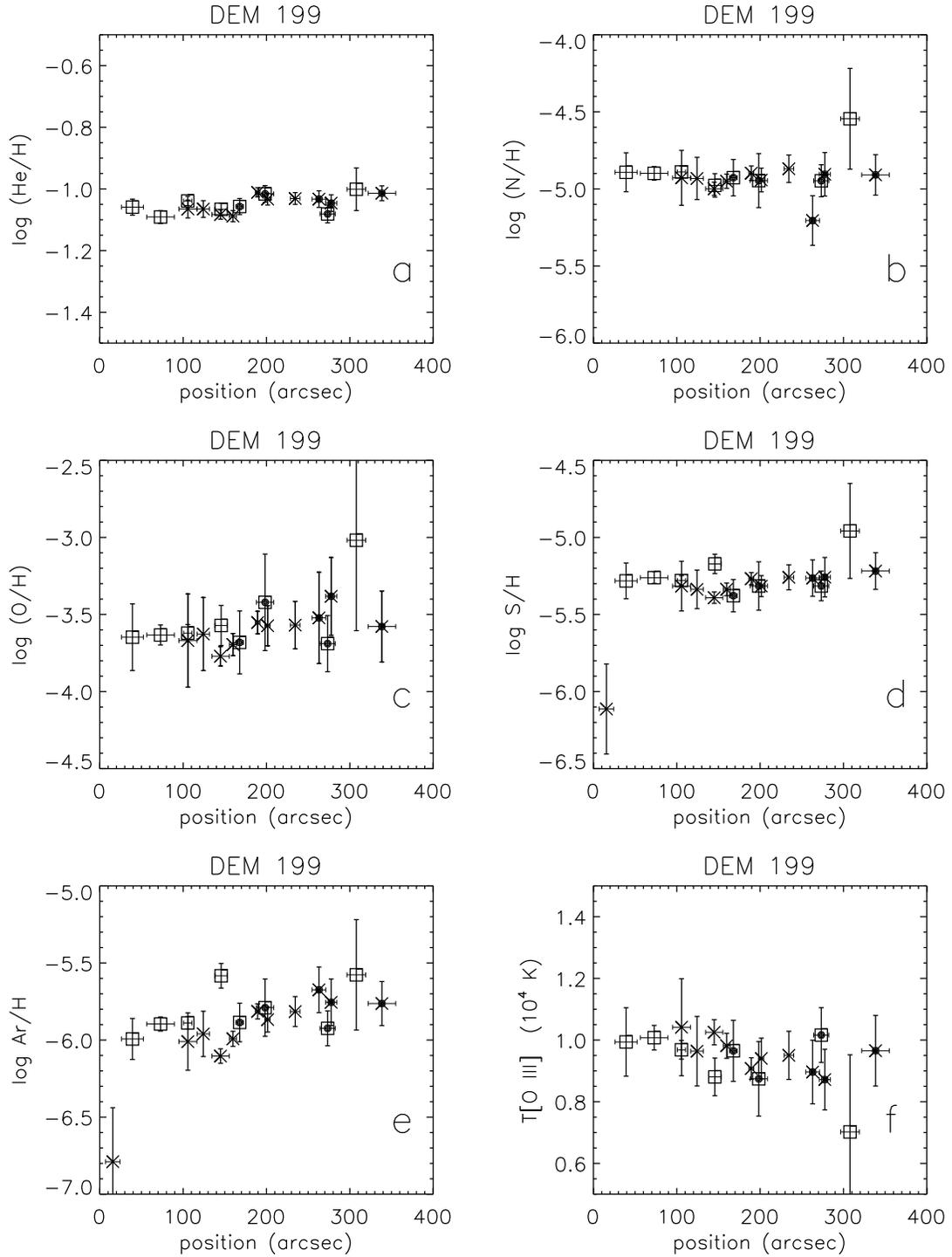}
\caption{Spatial variation in abundances and $T$\oiii\ along the slit
positions for DEM L199.  Light, horizontal lines show aperture sizes,
while vertical error bars indicate abundance uncertainties.  An
LMC distance of 50 kpc corresponds to 0.24 pc arcsec$^{-1}$.  
Apertures near WR stars are highlighted with solid dots.
\label{d199fig}}
\end{figure*}

\begin{figure*}
\epsscale{1.7}
\plotone{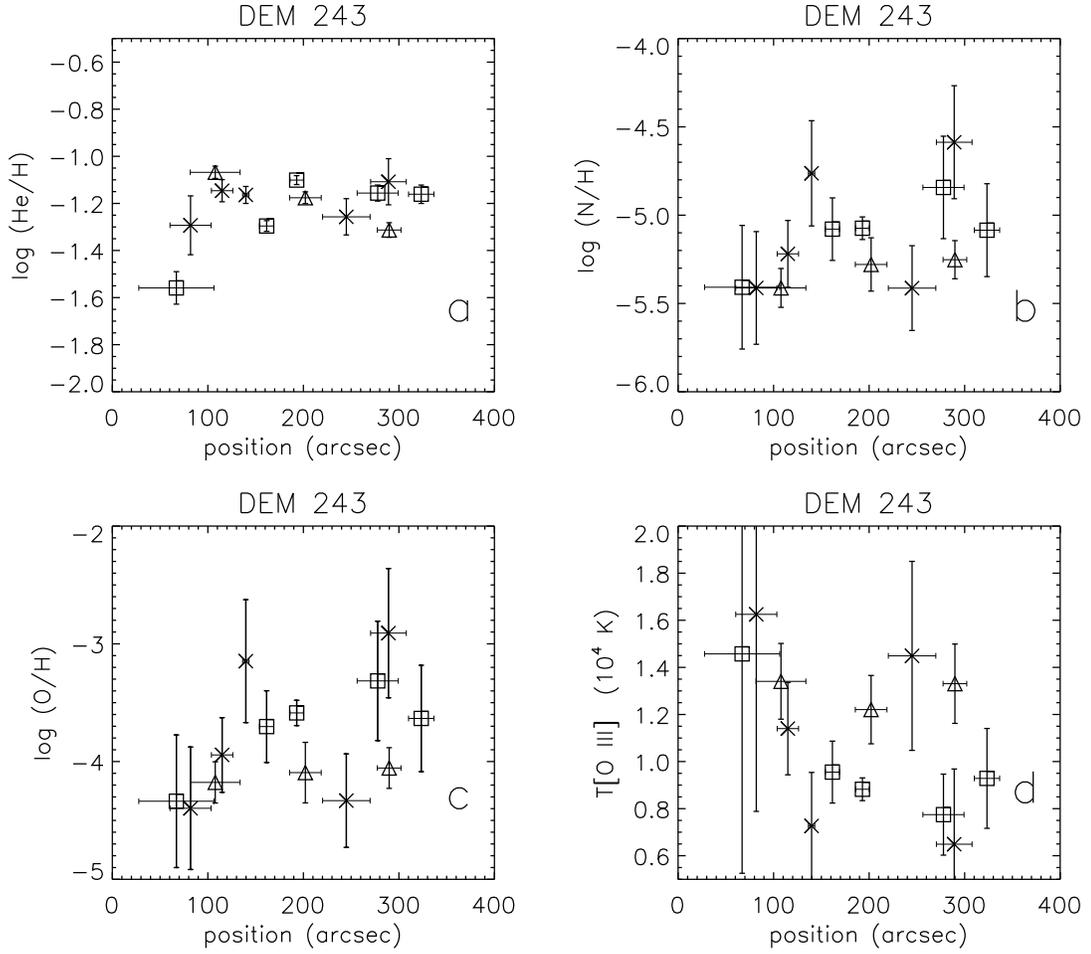}
\vspace*{-2in}
\caption{Spatial distribution for log(He/H), log(N/H), log(O/H), and
$T$\oiii\ for DEM L243.  The notation is as in Figure~\ref{d199fig},
but note the different scales.  Apertures affected by the SNR are
omitted.  Values of log(He/H) should be considered lower limits for
this object.
\label{d243fig}}
\end{figure*}

\begin{figure*}
\epsscale{1.7}
\plotone{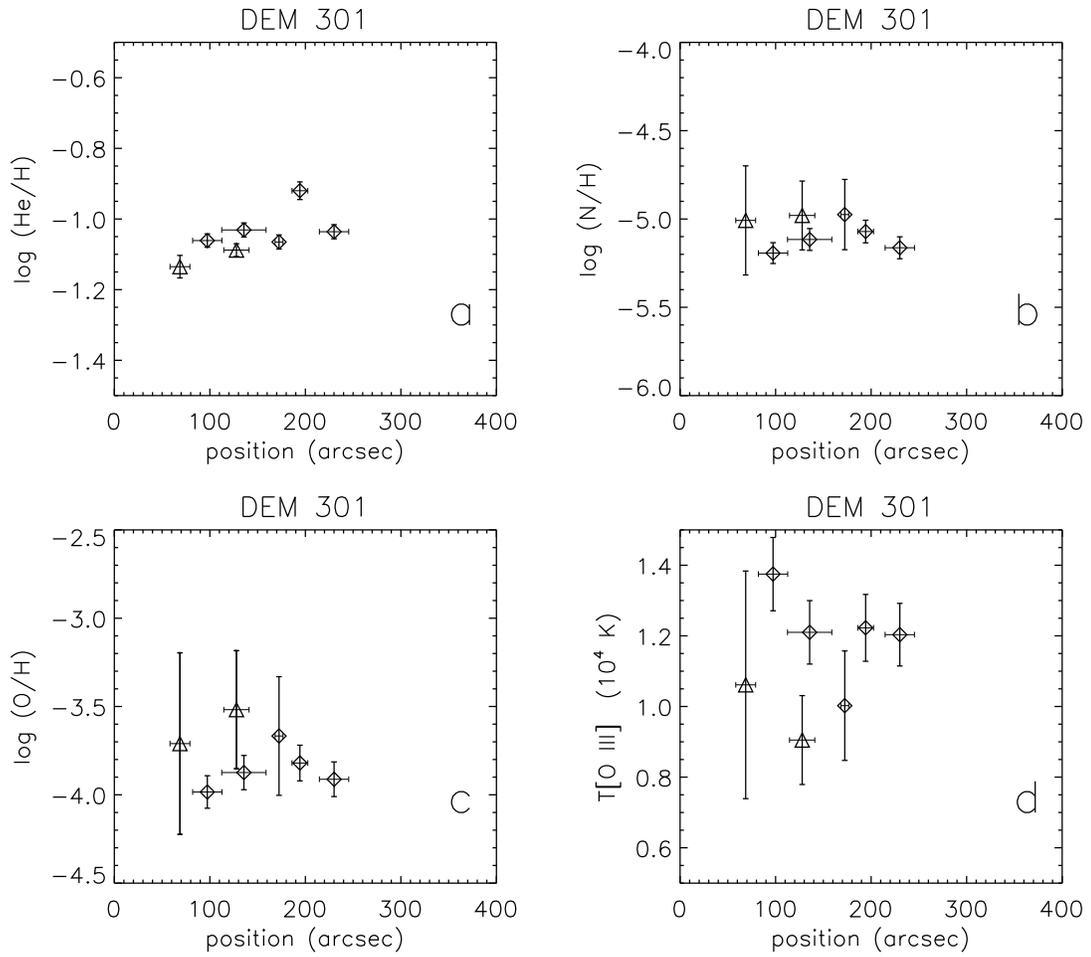}
\vspace*{-2in}
\caption{Same as Figure~\ref{d243fig}, for DEM L301.  Note the
ordinate scales are as in Figure~\ref{d199fig}.
\label{d301fig}}
\end{figure*}

\begin{figure*}
\epsscale{1.7}
\plotone{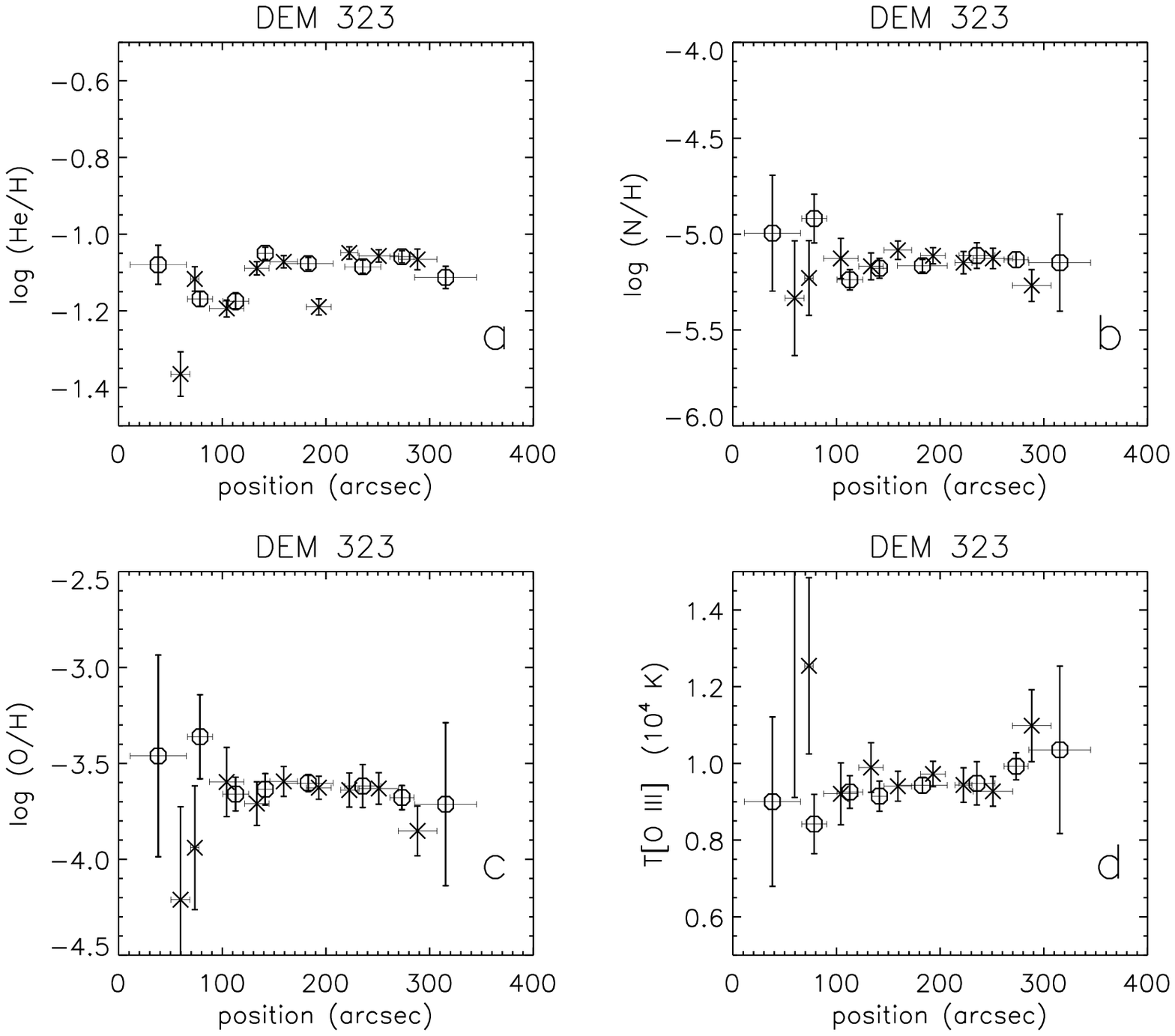}
\vspace*{-2in}
\caption{Same as the previous Figure, for DEM L323.
\label{d323fig}}
\end{figure*}

In Figure~\ref{d199fig}$a-e$, we show the spatial distribution of elemental
abundances across the sightlines for our stationary observations of
DEM L199.  Figure~\ref{d199fig}$f$ shows the corresponding
distribution in $T$\oiii\ measurements.  For DEM L243, DEM L301, and
DEM L323, we show in Figures~\ref{d243fig} -- \ref{d323fig} the
results for He, N, and O, as representative of the other elements.
These are shown respectively in panels $a,\ b$, and $c$; and
panels $d$ show corresponding measurements of
$T$\oiii.  The different symbols correspond to individual slit
positions as designated in Paper~I (Figures~6, 8, 12, and 13),
to aid cross-referencing.  Following our convention in that work, we
simply show the slit positions superposed, therefore these figures will not
necessarily show clean radial profiles across the nebulae, although
they do approximate this reasonably well.  This issue may be inspected in
Paper~I, along with actual slit positions.  The light,
horizontal lines indicate the 
spatial extent of each extracted aperture, while the vertical error
bars indicate uncertainties in the derived measurements.

\begin{figure*}
\epsscale{1.7}
\plotone{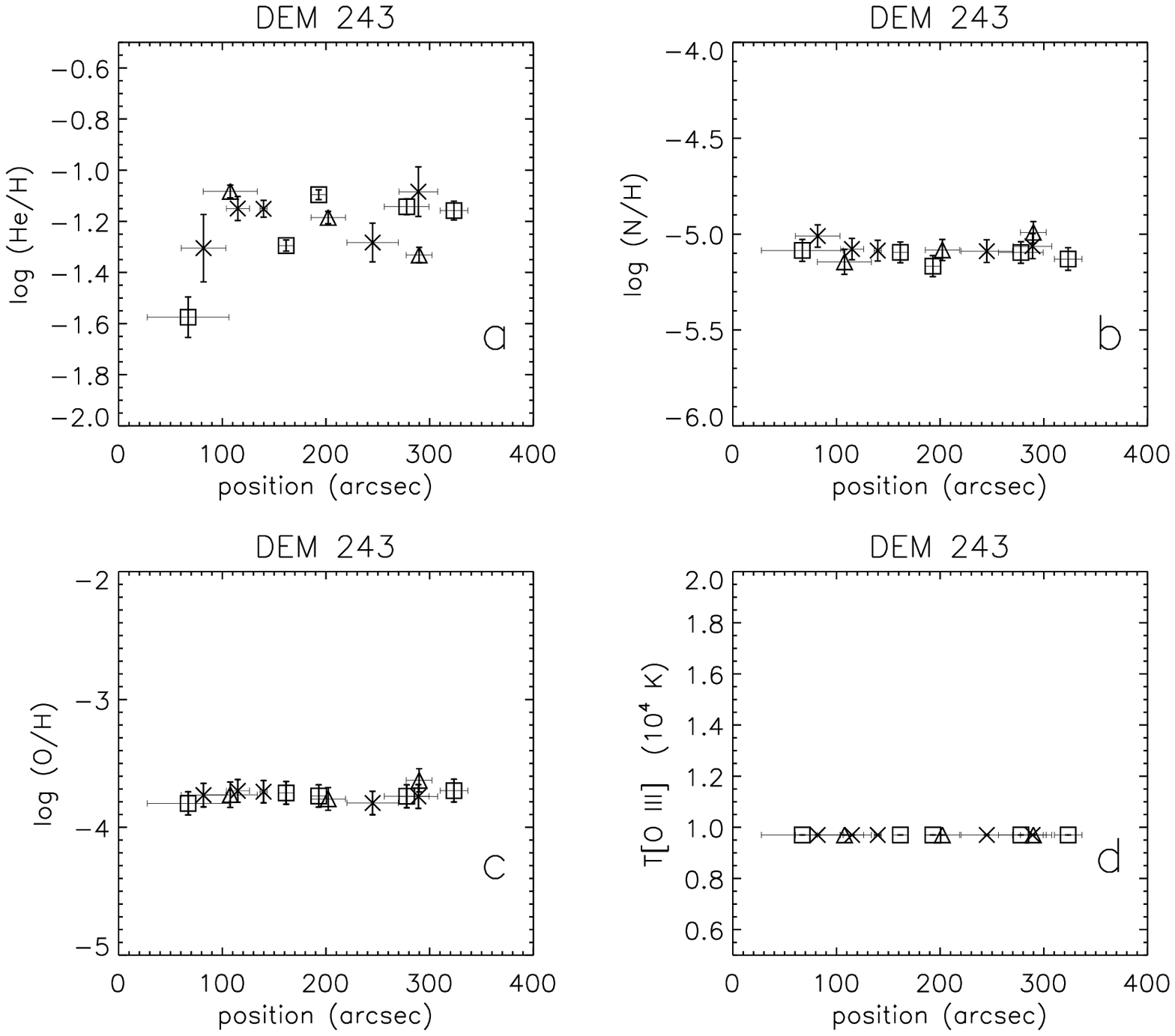}
\vspace*{-2in}
\caption{Data for DEM L243 as in Figure~\ref{d243fig}, but with \thi\
fixed at 9700~K for all included apertures.
\label{d243Tfig}}
\end{figure*}

In Figure~\ref{d243fig}, DEM L243 shows much larger scatter and
uncertainty among the individual apertures, which
can be attributed to the large measurement uncertainties for $T$\oiii\
(Figure~\ref{d243fig}$d$).  There is also an SNR either embedded or 
superimposed in the line of sight to this object, but we have not
included the affected apertures in our abundance estimates or in
Figure~\ref{d243fig}.  Assuming an essentially constant metallicity in
DEM L243, we recomputed the abundances by fixing \thi\ to be the mean
$T$\oiii, using the same weighting and omitting the same deviant
values as before, in computing the mean abundances.
Figure~\ref{d243Tfig} and Table~\ref{meandat} show the results using the 
new \thi$=$9700~K.  The dramatic reduction in scatter for
the heavy elements suggests that the original scatter
in Figure~\ref{d243fig} may indeed be caused by poor measurements in
$T$\oiii.  However, we emphasize a point by Mathis, Chu, \& Peterson
(1985), that true abundance fluctuations will induce corresponding
fluctuations in \te, owing to the more efficient cooling accompanying
higher O/H.  The reduced scatter with a fixed \te\ is therefore not a
strong demonstration of truly constant abundances.
To further test consistency with constant abundance
distributions, we computed the reduced-$\chi^2$ statistics for the
original distribution, yielding
1.4 and 1.8 for log(N/H) and log(O/H), respectively.  The probabilities
of obtaining these values for 12 degrees of freedom are 16\% and 4\%,
respectively.  These values therefore hint that the scatter may in
part be caused by real variations, although the significance is low.

In the event that the abundances for DEM L243 are essentially
constant, the mean values obtained with fixed $T$\oiii\ should give a
better estimate than the original values (Table~\ref{meandat}).
However, we caution that there is systematic uncertainty
introduced by the adopted value of \thi; comparison to the original
mean abundances in Table~\ref{meandat} suggests the uncertainty is
roughly 0.1 dex in the derived metallicities.  Since the recombination
lines used to determine log(He/H) are less temperature-sensitive than the
collisional metal lines, log(He/H) (Figure~\ref{d243Tfig}$a$) still
shows larger scatter than exhibited in the other objects.  However,
note that the 
magnitude of the variation is only $\sim0.3$ dex.  As discussed in
Paper~I, the \ion{He}{1}\ \lam5876/\Hb\ ratios suggest that He is not
uniformly fully ionized in DEM L243, therefore causing log(He/H) to be
underestimated in many apertures.  Our value for the He abundance is
therefore a lower limit in this object.  The upper envelope to the
distribution in Figure~\ref{d243Tfig}$a$ is around log(He/H) $\sim
-1.1$, which may be more indicative of the true He abundance.  This
value is more consistent with those for the other objects
(Table~\ref{meandat}). 

Thus, in Figures~\ref{d199fig}--\ref{d323fig} the
abundances appear to show no spatial variations within the measurement
uncertainties, with standard deviations typically around 0.10 -- 0.15 dex.
The uniformity of the abundance derivations is reassuring, and suggests
that our adopted ionic relations (\S 2.2) and description of the nebular
temperature structure (equation~\ref{garneq}) yield abundance
estimates with high accuracy.  The apparent success of these methods
is consistent with the finding by Mathis (1985) that ICFs for model
nebulae appear fairly robust between volume averaged and radially
averaged regions.  The uniformity of results for apertures tracing
different parts of the same nebula further suggests that spurious
``pencil-beam'' effects resulting from projection of radial \te\
gradients (Gruenwald \& Viegas 1992) are negligible for our objects.
This is consistent with the fact that sources in our sample have
$T$\oiii\ restricted to the approximate range 9500 -- 12,000 K, for
which the nebulae are expected to be relatively isothermal
(equation~\ref{garneq}; Figure~\ref{TO3O2}).  We emphasize that this
propitious circumstance may not apply to other \hii\ regions, that
may exhibit stronger \te\ gradients; as an example, Walter, Dufour, \&
Hester (1992) reported a significant \te\ gradient in the Orion
nebula, although their spatial scale of 0.5 pc is almost an order of
magnitude smaller than is relevant for our objects in the present study.
As in \S 2.1 above, we again emphasize the importance of choosing the
correct parameterization for the \te\ structure.

\subsubsection{Search for self-enrichment}

It is especially interesting to examine DEM L199 for possible
abundance variations that are introduced by the three WR
stars.  Two of these are binaries with a WN3 component (Breysacher
1981) and one is a WC4 + O6 V-III binary (Moffat {\etal}1990).  The
exact location of these stars with respect to the nebular gas
distribution and slit positions may be examined in Paper~I.  We emphasize
that DEM L199 is not a WR ejecta shell, but is a large, luminous
\hii\ complex whose dominant ionizing stars are early WR stars.
Walsh \& Roy (1989) and Kobulnicky {\etal}(1997) identified two
regions in the starburst galaxy NGC 5253 that appear to show enhanced
N, which is suggested to result from self-enrichment by WR stars.
Kobulnicky {\etal}(1997) point out that accompanying He
enrichment would be expected from the WR sources, and acknowledge that
the lack of He excess in these objects is puzzling.

DEM L199 has three WR stars within a large central cavity in a
luminous \hii\ region.  It would seem likely that, if self-enrichment
from WR stars can be seen, it should be detected in our observations.
In Figure~\ref{d199fig}, we indicate six apertures that are closest to
the stars Br~32 (WC4) and Br~33 (WN3), with solid dots.  These
apertures are identified in Paper~I as:  D199.205-15, 16, and 17; and
D199.205N120-17, 18, and 19.  The characteristic physical distances
from the WR stars are about 5 -- 15 pc.  Figure~\ref{d199fig} shows
that these points do not show any sign of abundance enhancements. 
Results from the full set of aperture measurements for DEM L199 are
consistent with uniform abundances at the 90\% confidence level, as
indicated by reduced-$\chi^2$ values of 0.59 and 0.60, for log(N/H)
and log(O/H) respectively, with 17 degrees of freedom. 
For He, in particular, the variation is constrained to be $\lesssim 0.1$ dex.
Although abundance enhancements are seen in WR ejecta nebulae, our
data suggest that self-enrichment by WR stars can be an extremely
subtle phenomenon.  It is possible that the processed material is
heated to coronal temperatures within the superbubble and is thus
optically undetectable.  Alternatively, the WR phase in these stars
may not yet have lasted long enough to produce significant enrichment in
the surrounding environment.

We were also interested to see whether the SNR in DEM L243 showed
evidence of abundance anomalies.  Abundance estimates from SNRs are
necessarily more uncertain, owing to the more complicated radiative
transfer, but meaningful estimates have been made by, e.g., Russell \&
Dopita (1990).  These are based on matching the emission-line spectra
with a shock code.  In Paper~I, we were able to find excellent
agreement between {\sc Mappings} shock models and the observed
emission from the shock-affected apertures, using the abundances
derived from the uncontaminated regions of the nebula listed in
Table~\ref{meandat}.  The SNR therefore shows no evidence of abundance
anomalies, although strong emission from photoionized gas in the same
line of sight prevents strong constraints.    

Likewise, the superbubble DEM L301 shows strong evidence of a recent SNR
impact (Chu \& Mac Low 1990; Oey 1996b; Paper~I), and thus could
conceivably show enrichment by massive star winds and supernova
ejecta.  However, its elemental abundances are often lower than for
the other objects in the sample (Table~\ref{meandat}).  We caution
that shock-excitation can significantly affect the abundance 
determinations for this object (Paper~I).  Peimbert, Sarmiento, \&
Fierro (1991) showed that contamination by shock activity can cause
abundances to be underestimated, especially for higher-ionization
species.  Our measurements for DEM L301 are consistent with this
behavior, although in \S 2.3.3 we find that the metallicities for DEM L243 
do not appear strongly affected by SNR contamination.

\subsubsection{Spatially integrated abundances \\ and LMC metallicities}

We also derived abundances for the scanned, spatially integrated apertures
using the same methods.  These are presented in Table~\ref{scandat},
along with mean LMC \hii\ region abundances from compilations by
Dufour (1984).  A more recent compilation by Garnett (1999) shows
essentially the same values.  Within the uncertainties, the mean
abundances for each object from Table~\ref{meandat} agree with the
determinations from the spatially integrated observations
(Table~\ref{scandat}), although the offsets appear to be
systematic across all elements for each object.  This again points to
uncertainties in the \thi\ determination, which can result from simple
measurement errors, or factors related to the spatial integration of
the line emission.  We regard the values obtained from the mean of the
spatially resolved data (Table~\ref{meandat}) to be more reliable than
the single observations from the integrated data.  However, we caution
that the ionic relations in \S 2.2 are based on spatially
integrated models and observations, and this could introduce
systematic variations between the spatially resolved and scanned
data.  But taking the derived values and uncertainties at face value,
the mean abundances of the stationary apertures should be somewhat
more reliable.

We include in Table~\ref{scandat} the abundances for DEM L243 derived
from both the total, scanned observation and the scanned observation
with the SNR-contaminated region subtracted.  Interestingly, there is
no significant difference between these, although the data including
the SNR do show the expected decrease in computed abundances (Peimbert
{\etal}1991).  Thus, while DEM L301 showed suspiciously low
metallicity measurements attributable to effects of shock emission,
DEM L243 is an example where the SNR 
is not a significant factor.  In Paper~I, we also found that the two
shock-affected objects exhibit different behavior in their line
emission with respect to the photoionized regions, thereby
demonstrating how shocks contribute
in different ways to the spectra of host \hii\ regions,
depending on shock velocity and environment. 

We find a tendency for our measurements to be $\sim 0.2$ dex lower
than the mean LMC metal abundances compiled by Dufour (1984).
One of the probable causes is our adopted 
temperature structure (equation~\ref{garneq}), which varies slightly
from those used by others.  For example, we find that our mean
abundances for DEM L243 would increase by about 0.1 dex if we adopted
the relation of Campbell {\etal}(1986) at all values of \thi.

Our data are generally consistent with there being no abundance
variations between the four different \hii\ regions.  It is
interesting to note that DEM L199 is close to the LMC bar, about 1 kpc
from the center of the galaxy; and DEM L243 is situated in the
northern outskirts of the LMC-4 supergiant shell, at a galactocentric
radius of $\sim$3.5 kpc.  Pagel {\etal}(1978) have suggested that the
LMC \hii\ regions possibly exhibit a slight abundance gradient.  This
has not been further examined, nor has a gradient been
detected in the cluster population (Olszewski {\etal}1991).  In our
data, it is suggestive that DEM L199 and DEM L243 delineate the
extremes of any interpreted variation among our four objects.  The
difference in metallicity is consistent with the small gradient suggested
by Pagel {\etal}(1978).  

\section{Semi-empirical bright-line methods}

We turn now to examining more indirect emission-line diagnostics of
metal abundances.  In situations where \te\ cannot be adequately constrained by
observation, it is common practice to estimate the metal abundances
using the semi-empirical, ``bright-line'' abundance parameters.  Here
we examine the performance of these parameters in light of our
detailed nebular data and highly-constrained, tailored photoionization
models from Paper~I.

We also compute model tracks of the abundance parameters, using {\sc
Mappings II} with generalized nebular parameters.
These incorporate the stellar energy distribution of CoStar
model C2 (Schaerer \& de Koter 1997), which corresponds to an O6 -- O7
stellar effective temperature.  We assume an inner radius to the gas
distribution of 0.4$R_{\rm S}$, and gas density $n=10\cc$.  In the
figures that follow, the dashed, solid, and dotted lines correspond to
the volume-averaged $\log U = -2,\ -3,$ and $-4$, respectively, which is
equivalent to changing the total ionizing photon emission rate or gas
filling factor.

The grid of models is computed with $Z = 0.05,$ 0.1, 0.3, 0.5, 1.0, and
2.0 times $\Zsol$.  We included the elements (He, C, N, O,
Ne, Mg, Al, Si, S, Ar, Ca, and Fe) with $\Zsol = (-1.01, -3.44, -3.95,
-3.07, -3.91, -4.42,$ $-5.53, -4.45, -4.79, -5.44, -5.88$, and --4.96),
respectively.  We largely follow McGaugh (1991) in scaling the abundances of
individual elements with respect to O.  For He and N, we use the relations
given by McGaugh, but scaled to match the Anders \& Grevesse (1989)
values for $\Zsol$:
\begin{equation}
\rm He/H = 0.0850 + 15 (O/H)
\end{equation}
and
\begin{equation}
\rm \log (N/H) = 1.5\ \log(O/H) + 0.66 \quad .
\end{equation}
For C and Fe, we adopt McGaugh's relations directly (his equations~10
and 11).  The remainder of
the elements are fixed in their proportion to O at $\Zsol$, as given
by Anders \& Grevesse.  Models for $\Zsol$ and $2\Zsol$ use the 
CoStar C2 atmosphere at solar metallicity, while the rest use the 
corresponding SMC metallicity model; we find that the stellar
metallicity is unimportant for these $U$-tracks.

\subsection{$R23$}

\begin{figure*}
\vspace*{-1.0in}
\epsscale{2.0}
\plotone{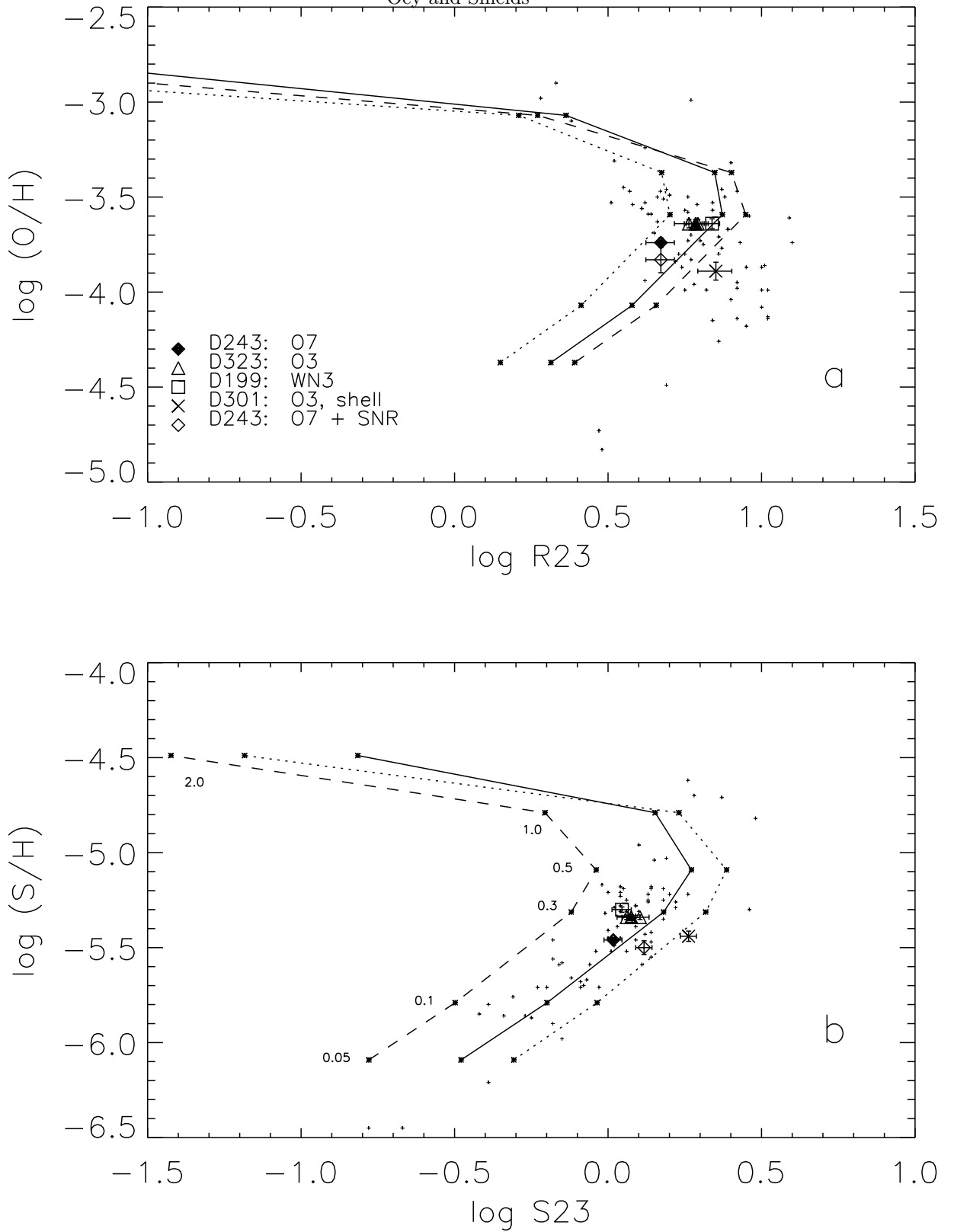}
\caption{Abundance parameters $\log R23$ (panel $a$) and $\log S23$
(panel $b$) vs. log(O/H) and log(S/H), respectively.
The tracks indicate photoionization models for 0.05, 0.1, 0.3, 0.5,
1.0, and 2.0$\Zsol$, as labeled in panel $b$.  The
dashed, solid, and dotted linetypes correspond to $\log U = -2,\ -3,$
and $-4$, respectively.  Our observations
are indicated by the symbols shown in the key; the solid triangle
corresponds to observations for DEM L323 scanned over the entire
object, while the open triangles represent partial scans (see
Paper~I).  The small plus signs show data compiled by DPM, for objects
whose abundances are determined using direct measurements of $T_e$.
\label{abdparam}}
\end{figure*}

The O abundance parameter $R23$ (Pagel {\etal}1979; equation~\ref{eqR23}
above), has been extensively used and empirically calibrated several
times, by McGaugh (1991), Skillman (1989), and Dopita \&
Evans (1986), among others.  In
Figure~\ref{abdparam}$a$, we show \logOH\ vs. $\log R23$, with the tracks
showing results from generalized photoionization models described above.
Our assumptions differ somewhat from those of previous studies, in
particular with the assumption of a fairly hollow morphology and new
stellar atmosphere models.  Most authors (e.g., McGaugh 1991; Dopita
\& Evans 1986) also assume some anticorrelation between the 
characteristic \tstar\ or $U$, and $Z$, in their adopted calibration at high
abundance.  Our tracks do not assume this anticorrelation.
Despite these differences, our tracks remain similar to those of
previous authors, although our models show a slight offset to higher
log(O/H) (see Kobulnicky {\etal}1999 and McGaugh 1991 for 
comparisons of $R23$ calibrations).

We also plot in Figure~\ref{abdparam}$a$ the spatially integrated data for
our objects, using the abundances derived from the means of our
resolved apertures (Table~\ref{meandat}).  
The values of $R23$ are computed in Paper~I and
shown here in Table~\ref{tababdpar}.  We caution that the 
nebular fractional area included in the spatial scans varies among the
four objects, and we refer the reader to Paper~I for the precise details.
The three spatial scans
of the spherical object DEM~L323 (triangles) should give an indication of the
degree to which subsampling is representative of the total spatial
scan (solid triangle).  The three measurements of $R23$ are in
excellent agreement, which is consistent with our finding in Paper~I
that this index is robust to spatial variations.
For DEM~L243, we show $R23$ derived from the spectrum
with the SNR-contaminated region subtracted (solid diamond);
and also that from the total integrated region including the SNR
(open diamond).  The square and cross show DEM~L199 and DEM~L301,
respectively. 

Our data points are generally well-behaved with respect to the model
tracks in Figure~\ref{abdparam}$a$.  While we found excellent agreement
between the observed emission-line spectra and our tailored photoionization
models in Paper~I, we see in Figure~\ref{abdparam}$a$ that most of the
objects fall in their expected location with respect to the more
generalized model tracks.  DEM L323, DEM~L243, and DEM L199 fall
between tracks of $\log U = -3$ and $-4$, with DEM L199 showing the
highest value of $U$, as expected in this high-excitation object.  The
one anomalous point is DEM L301 (cross), which we found in Paper~I to have
an unusual combination of excitation mechanisms.  We concluded that
this object, which has an extreme shell morphology, is most likely
ionized by a combination of density-bounded photoionization plus
shocks.  While our tailored model for this object reproduced the
observed $R23$ well, it is apparent that its value is anomalously
high with respect to the tracks in Figure~\ref{abdparam}$a$.  Ironically,
the offset is in the sense of a higher ionization parameter, although the
object in fact has a much lower $U$ than the others in the sample.
The larger value of $R23$ is probably caused by the enhanced emission
contributed by the shock activity.  DEM L301 and DEM L199 have similar
values of $R23$, which, lacking any additional information for these
objects, would imply similar abundances; we see that in fact this would
overestimate \logOH\ for DEM L301 by about 0.3 dex, taking the
measured \logOH\ at face value.

We also overplot with small plus signs in Figure~\ref{abdparam}$a$ the
data compiled by D\'\i az \& P\'erez-Montero (2000).
These represent data from the literature (their Table~2) for which
abundances were derived from a direct measurement of $T$\oiii.
Although the points for our data from Paper~I are consistent with the
model tracks, it is apparent that the photoionization models in
Figure~\ref{abdparam}$a$ do not track well the locus of the larger
dataset.  It is important to note that adopting softer
stellar atmospheres can improve the correspondence slightly, since
this would offset the tracks to slightly lower log(O/H).  We refer
the reader to McGaugh (1991) to evaluate the consequences of the stellar
effective temperature.  The atmospheres adopted here (CoStar C2)
correspond to O6 -- O7 stars, which are already cooler than
most of our LMC objects, and only relatively small changes result if O3 -- O4
atmospheres (CoStar E2) are used instead.  The discrepancy between models
and data has always been a difficulty in the use of $R23$, and
therefore empirical calibrations of this parameter have been crucial
for its successful use.


\subsection{$S23$}

As mentioned earlier, a parameter similar to $R23$ has been introduced
for S by V\'\i lchez \& Esteban (1996) and Christensen {\etal}(1997),
which was further explored by D\'\i az \& P\'erez-Montero
(2000, hereafter DPM).  It is important to note that this parameter, $S23$
(equation~\ref{eqS23}), is {\it not} strictly analogous to $R23$.
While O and S have homologous energy levels, the
ionization potentials (IP) for their respective ions are different.  In
particular, it is important to note that the IP required to reach \sp3\
(34.83 eV) is virtually identical to that necessary for \opp\
(35.12 eV).  Therefore, although S$^+$ and \spp\ are indeed the
dominant ions for S, for typical \hii\ regions,
there is likely to be non-negligible \sp3, which is ionized by the
same radiation that produces \opp.  Although Christensen {\etal}(1997)
pointed out that the ionization fraction of \sp3\ is relatively small,
typically $\lesssim 0.2$, we show below that it nevertheless 
significantly affects the ionization balance
of S$^+$ and \spp, and consequently, the value of $S23$.

Figure~\ref{abdparam}$b$ is similar to panel $a$, now showing $\log$
(S/H) vs $\log S23$ (Table~\ref{tababdpar}).  The model line types and
data symbols are the 
same as before.  It is immediately apparent that, contrary to earlier
claims in the literature, $S23$ is {\it more} sensitive to the
ionization parameter than $R23$.  The change in $\log S23$ between
the model tracks, varying $\log U$ from --2 to --4, is almost 0.5 dex,
whereas it is less than 0.3 dex for $R23$.  The greater
$U$-sensitivity of $S23$ is caused by the ``missing'' contribution of
\ion{S}{4}.  Figure~\ref{abdparam}$b$ shows that the models with high
$U$ show lower $\log S23$, the opposite pattern to $R23$.  This is
consistent with the ionization behavior of S, since a larger population of
\sp3\ is expected at higher $U$.

As suggested by DPM, the lower-metallicity branch of $S23$ does span a
larger range in values than $R23$, for a given range of $Z$
(Figure~\ref{abdparam}).  However, the location of the inflection at
the maximum $S23$ is at only a slightly higher $Z$ than that for
$R23$.  Figure~\ref{abdparam} shows that our models for $\log R23$ have
a maximum close to $Z=0.3\ \Zsol$, and those for $\log S23$ have a
maximum around $Z=0.5\ \Zsol$.  Thus, there is only $\sim 0.2$ dex
extension in the dynamic range of $Z$ in the use of $S23$.
Nevertheless, since so many of the observed objects in the literature
have abundances around $0.3 - 0.5\Zsol$, this augmentation makes a substantial
difference in evaluating abundances.  As is dramatically shown in
Figure~1 of DPM, $S23$ empirically shows  
an evidently monotonic increase as a function of $Z$, in contrast
to $R23$, which shows a distinct double-valued structure in $Z$.

Our data points are again reasonably consistent with the models in
Figure~\ref{abdparam}, although they now perhaps show a tendency to
fall between the $\log U=-2$ and $-3$ models, rather than --3 to --4 for
Figure~\ref{abdparam}$a$.  The tailored photoionization models and
spatially-resolved data in Paper~I showed similar minor
discrepancies.  The data points for the shock-affected regions,
DEM L301 (cross) and the SNR-contaminated observation
for DEM L243 (open diamond) are offset to higher $\log S23$.
This suggests that $S23$ is increased by the presence of shock
excitation, similar to the behavior of $R23$ found in the previous section.

We also see in Figure~\ref{abdparam}$b$ that the photoionization
models do track the data well for 
$S23$, and much better than for $R23$ in panel $a$.  However, it is
also apparent that the points with the highest values of $S23$ fall
outside the model tracks.  In formulating a calibration for $S23$, we
would therefore recommend that these values be excluded, and that the
empirical calibration derived by DPM should not be used for $Z \gtrsim
0.5\Zsol$.

\subsection{$S234$}

As discussed in the previous section, the population of 
\sp3, which is ionized by the same radiation as that ionizing \opp,
is not sampled by the $S23$ parameter.  While the ionization fraction
of \sp3\ is relatively small, we have seen in the previous section
that it significantly compromises the utility of $S23$ as an abundance
diagnostic.  We therefore suggest that the parameter,
\begin{equation}\label{eqS234}
S234 \equiv \rm\frac{[S\thinspace II]\lambda6724 + 
	[S\thinspace III]\lambda\lambda9069,9532 +
	[S\thinspace IV]\lambda10.5\mu}{\hb}
\end{equation}  
is a better abundance indicator than $S23$.  In the same way that
$R23$ samples all significant ions of O, $S234$ more completely
samples the significant ions of S.  Note that any population of
S$^{+4}$ (IP 47.30 eV) will be an insignificant fraction 
of the total S ions, for massive star sources:  S$^{+4}$/S $\lesssim
0.02$ even for an object ionized by a WR star where $\rm He^{++}/He = 0.4$.
The principal difficulty with $S234$ is the inclusion of the mid-IR line
[\ion{S}{4}]$\lambda10.5\mu$, which is not readily observable with
standard ground-based instrumentation.  However, since the IP
necessary to produce \sp3\ is virtually the same as that for \opp,
it is possible to estimate the abundance of \sp3 based on that for
\opp.  This was demonstrated earlier by Mathis (1982) and
Dennefeld \& Stasi\'nska (1983).  We present this approach here.

\begin{figure*}
\epsscale{1.7}
\plotone{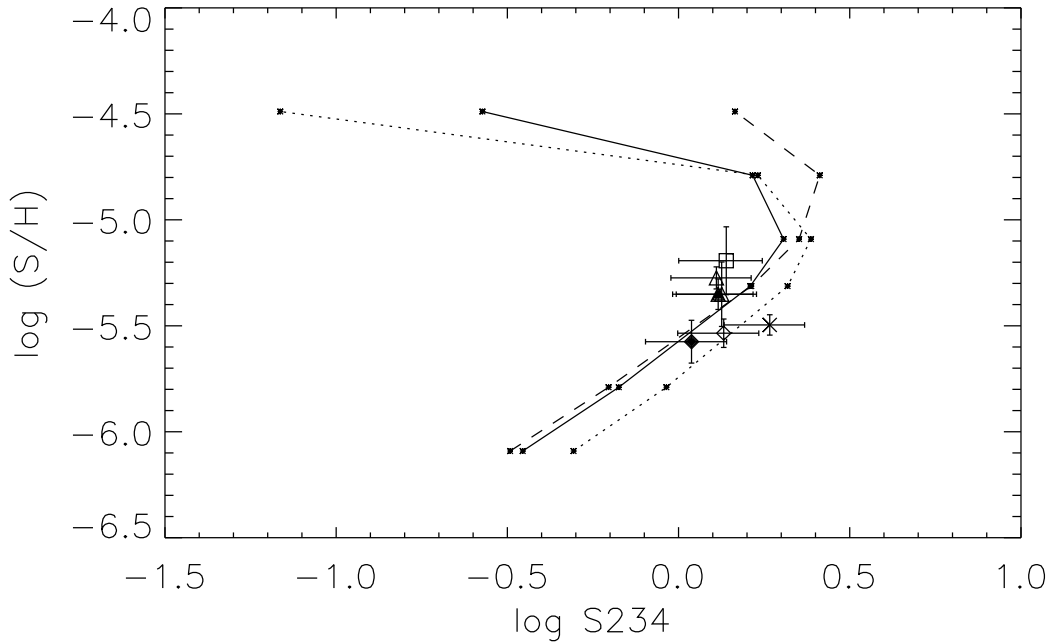}
\vspace*{-3.7in}
\caption{Photoionization models and data for log(S/H) vs. $S234$.  As
before, the tracks are computed for metallicities 0.05, 
0.1, 0.3, 0.5, 1.0, and 2.0$\Zsol$; and dashed, solid, and dotted
linetypes correspond to $\log U = -2,\ -3,$ and $-4$, respectively.
Observed values for $S234$ are computed as described in the 
text, and are indicated by the symbols as in Figure~\ref{abdparam}.   
\label{S234}}
\end{figure*}

Figure~\ref{S234} presents $\log$ (S/H) vs. $S234$, on the same scale
as that for $S23$ in Figure~\ref{abdparam}$b$.  The line types and
symbols are the same as before.  We see that $S234$ is 
dramatically less sensitive to the ionization parameter, owing to the
inclusion of the \ion{S}{4} indicator.  Indeed, the models for
$S234$ are even less sensitive to $U$ than is $R23$
(Figure~\ref{abdparam}$a$), for $Z \lesssim\Zsol$.  

\begin{figure*}
\epsscale{1.7}
\plotone{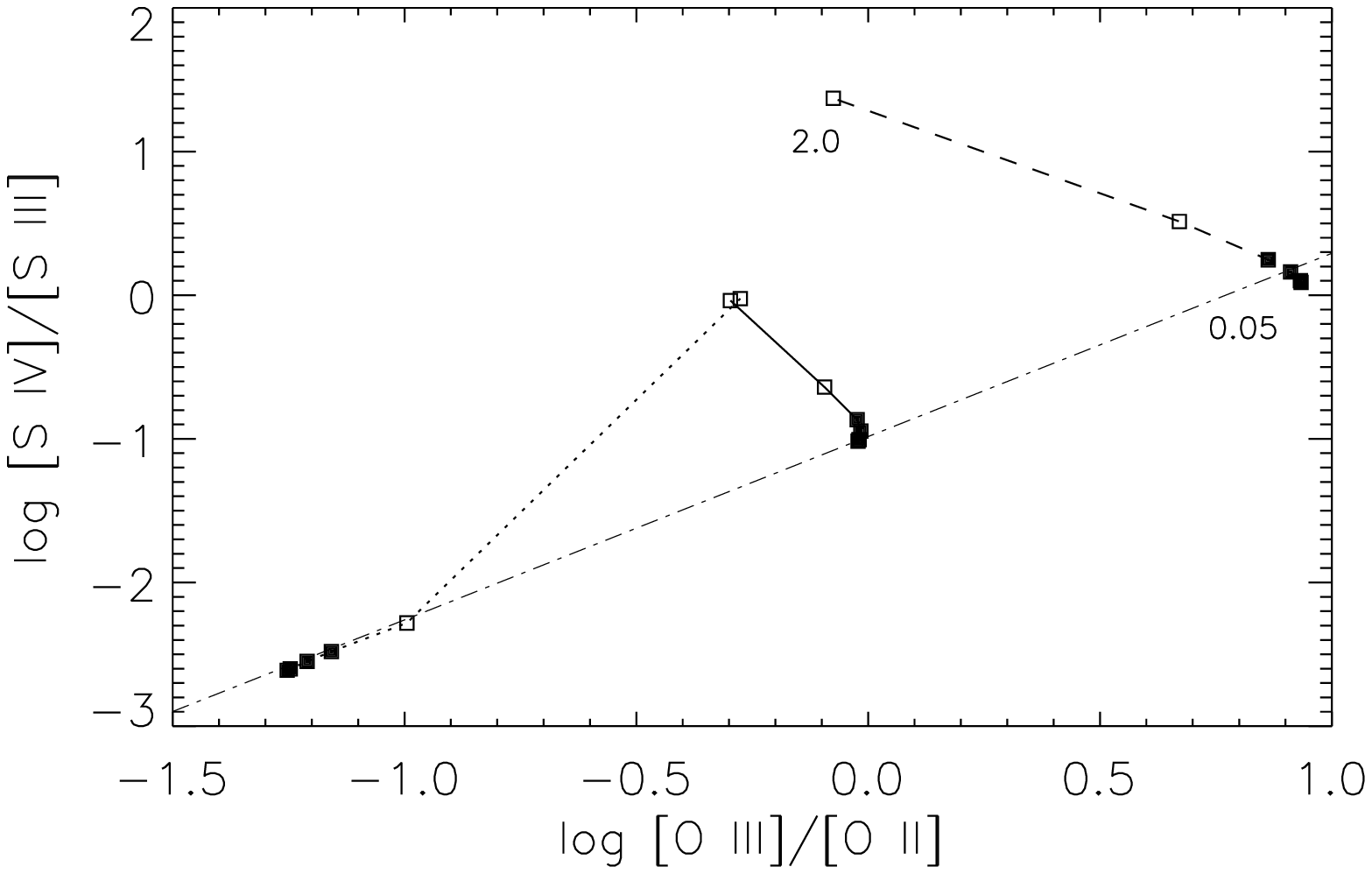}
\vspace*{-3.7in}
\caption{Models of log(\siv\lam10.5$\mu$/\siii\lam\lam9069,9532) vs.
log(\oiii\lam\lam4959,5007/\oii\lam3727).  These models employ the
Costar E2 stellar atmosphere model.
The line types are as before, and the metallicity extremes are
indicated in solar units for the $U = -2$ track (models for 0.05 and
0.1$\Zsol$ are virtually degenerate and indistinguishable).
Equation~\ref{eqS4O3} (dot-dashed line) is fitted from the solid
points ($Z\leq 0.5\Zsol$).  
\label{S4O3}}
\end{figure*}

It would therefore be desireable to estimate the intensity of \sivlam\
from that of the O ionization indicator, \oiii/\oii.  
Figure~\ref{S4O3} shows \break log(\sivlam/\siii\lam\lam9069,9532) vs.
\hfill\break log(\oiii\lam\lam4959,5007/\oii\lam3727) for models with an E2
CoStar atmosphere, and line types as before.  This atmosphere
corresponds to an O3 -- O4 stellar type, and we prefer this model in
examining the relation between \ion{S}{4} and other ions since it is
more relevant in environments with harder ionizing fields.
We see that for $Z\leq
0.5\Zsol$ (solid points), the relation 
between these ratios is essentially a simple power law.  For these
points, we fit:
\begin{equation}\label{eqS4O3}
{\log{\frac{\rm [S\thinspace IV]\lambda10.5\mu}
	{\rm [S\thinspace III]\lambda\lambda9069,9532}} } = \break \\
	{-0.984 + 1.276\ \log{\frac{\rm [O\thinspace III]\lambda\lambda4959,5007}
	{\rm [O\thinspace II]\lambda3727}} } \quad ,
\end{equation}
which is shown by the dot-dashed line in Figure~\ref{S4O3}.

With observations of \siii, this relation allows an estimate of the
\siv\ intensity, which can then be used to compute $S234$.
We note that the adoption of the cooler C2 CoStar atmospheres, used in our
other photoionization models, would result in a difference of less than
0.02 and 0.01 in the fitted intercept and slope, respectively.
Furthermore, the correction for \siv\ will only be significant for
moderate to high $U$ and/or high $Z$.  As a test of
equation~\ref{eqS4O3}, we use mid-infrared and optical line
observations of the Orion nebula by Lester, Dinerstein, \& Rank
(1979).  For their measurements,
equation~\ref{eqS4O3} predicts a volume emissivity for
\siv\lam10.5$\mu$ of $6.5\pm 2.7\times 10^{-21}\ \rm erg\ cm^{-6}\ s^{-1}$,
which agrees within measurement uncertainties with the observed value
of $9.0\pm 2.7\times 10^{-21}\ \rm erg\ cm^{-6}\ s^{-1}$.  Considering
the extremely narrow, 10$\arcsec$ line of sight
on the Orion nebula, and much higher density ($10^4\ \cc$) than
considered for our purposes, this agreement is highly encouraging.

It is thus relatively simple to convert from $S23$ into $S234$ and
thereby almost eliminate the sensitivity to $U$.
We used this method of estimating \siv\ to compute $S234$ for our
spatially-integrated observations (Table~\ref{tababdpar}), which are
plotted in Figure~\ref{S234}, using the same 
symbols as before.  In the errors for $S234$, we include in quadrature
an uncertainty of 25\% for the uncertainty of \siv\ from
equation~\ref{eqS4O3}.  We again have excellent 
agreement with the models.  It is clear that $Z$ can be estimated
with greater confidence based on $S234$ than $S23$ at these
metallicities, because the large spread in the models for $S23$
(Figure~\ref{abdparam}$b$) has been vastly reduced for $S234$. 

\begin{figure*}
\epsscale{1.7}
\plotone{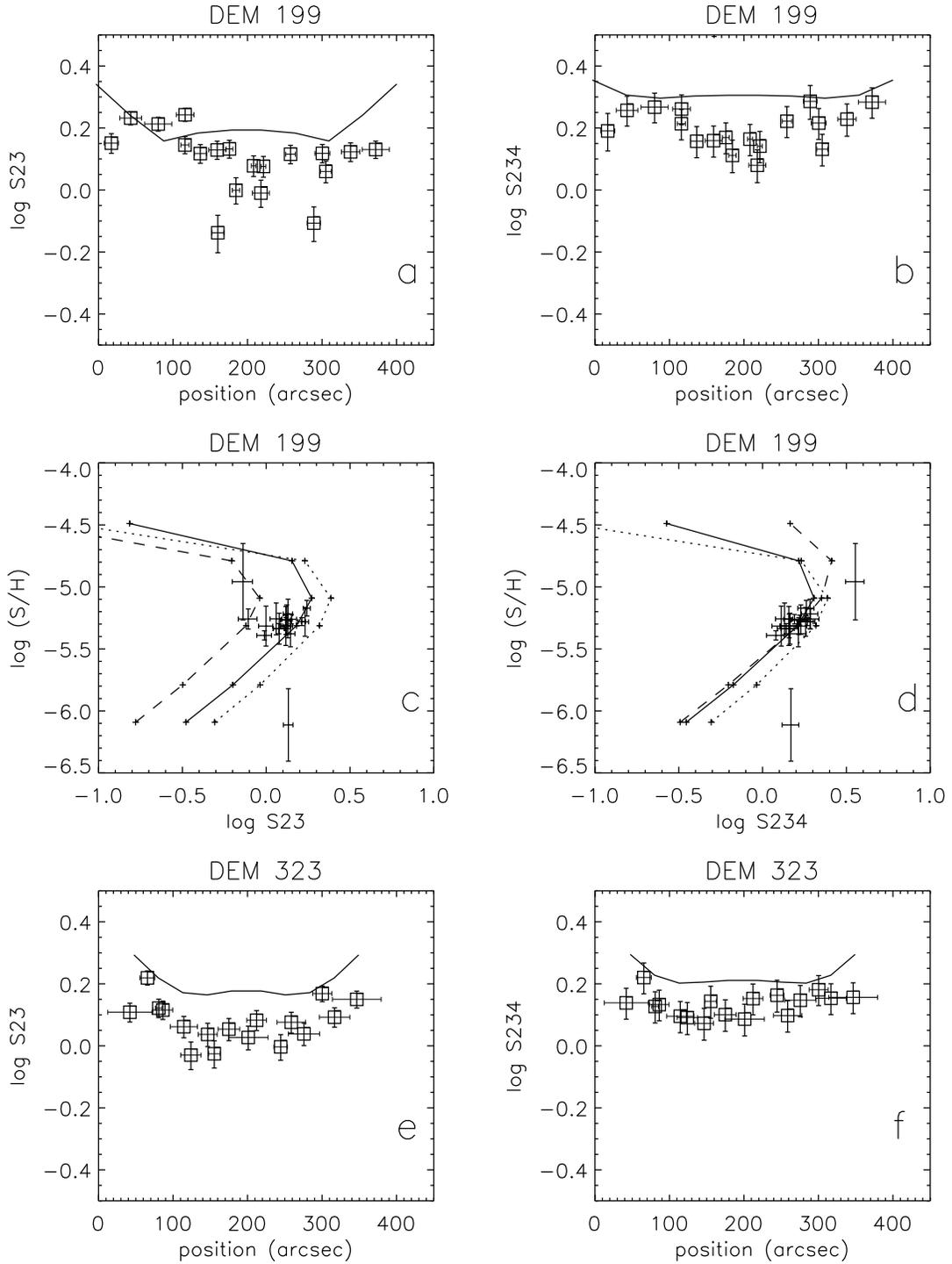}
\caption{Spatial variation of $S23$ and $S234$ in DEM L199 (panels $a$
and $b$) and DEM L323 (panels $e$ and $f$).  The light, horizontal
bars indicate the spatial extent of the apertures for these data.
Panels $c$ and $d$ show the same data for DEM L199 superimposed on the
models, with linetypes as in Figure~\ref{abdparam}.  
\label{S234spat}}
\end{figure*}

Similarly, we show in Figure~\ref{S234spat} that spatial variations
are also reduced from $S23$ to $S234$.  Figures~\ref{S234spat}$a$ and
$b$ show our spatially resolved observations of DEM L199 for $S23$ and
$S234$.  The solid line indicates the tailored model for this object,
using the early WR model of Schmutz et al. (1992; see Paper~I),
central hole radius of 0.5$R_{\rm S}$, and gas density $n=100\ \cc$.
While Figure~\ref{S234spat}$a$ shows a large spatial variation of
$\gtrsim 0.4$ dex for $S23$, we see in Figure~\ref{S234spat}$b$ that
the variation in $S234$ is reduced by about a factor of 2 in the
logarithm.  Figures 
~\ref{S234spat}$e-f$ show the same behavior for DEM L323.  The solid
line again represents the corresponding tailored model, with an O3 --
O4 stellar atmosphere (Costar E2), central
hole radius of 0.4$R_{\rm S}$, and gas density $n=10\ \cc$.  In
Figure~\ref{S234spat}$c-d$, we show the spatially resolved data for
DEM L199 superimposed on the model tracks for $S23$ and $S234$,
respectively.  The reduced scatter in $S234$ against the models again
demonstrates the improved constraints in estimating log(S/H), compared
to $S23$.

\subsection{Calibrations}

\begin{figure*}
\epsscale{1.7}
\plotone{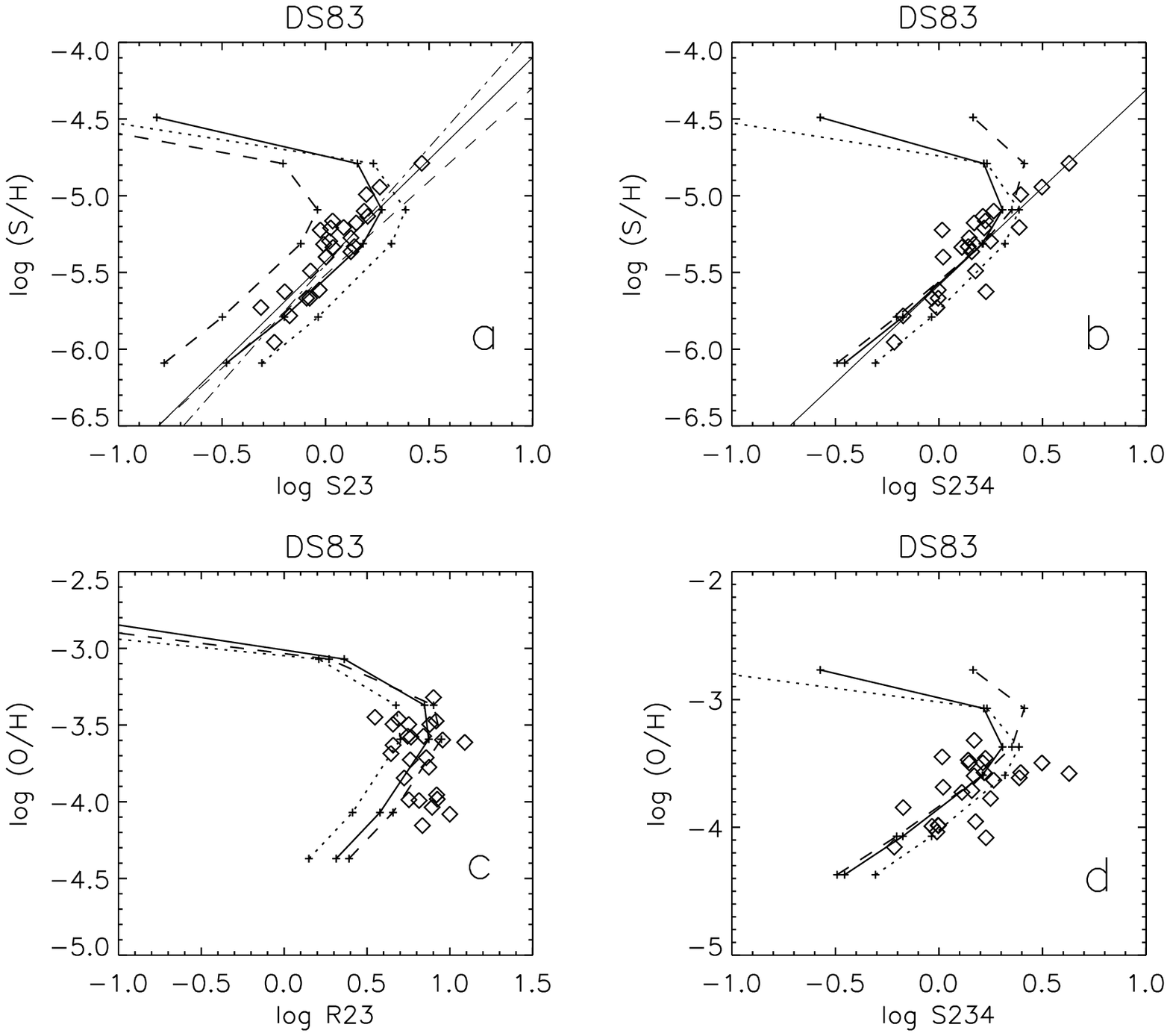}
\vspace*{-2in}
\caption{S abundance vs. diagnostics $S23$ (panel $a$) and $S234$
(panel $b$), and O abundance vs. $R23$ (panel $c$) and $S234$ (panel
$d$).  The models are the same as in Figures~\ref{abdparam} and
\ref{S234}.  Data points are from Dennefeld \& Stasi\'nska (1983),
with $S234$ computed as described in the text.  The lighter, solid lines
show a fit to our models (see text), while the light, dot-dashed and
dashed lines show the DPM and Christensen {\etal}(1997) empirical
calibrations, respectively. 
\label{DSfig}}
\end{figure*}

In Figure~\ref{DSfig}$a$ and $b$ we show the models for $\log ({\rm
S/H})$ as a function of $S23$ and $S234$, overplotted with the
Galactic and LMC data presented by Dennefeld \& Stasi\'nska (1983).
Their S abundances are computed from measurements of $T$\oiii\ and
observations of \siii\lam\lam9069,9532.  We compute $S234$ from this
dataset with the aid of equation~\ref{eqS4O3}, as described above.  
Figure~\ref{DSfig} shows that the data present a well-defined
sequence in both $S23$ and $S234$.  It is especially encouraging that
the locus of the models is in excellent agreement with that of the
data, in contrast with the situation for $R23$, as we saw above in
Figure~\ref{abdparam}$a$.  We replot the $R23$ models with the Dennefeld
\& Stasi\'nska data in Figure~\ref{DSfig}$c$, again suggesting the same
discrepancy seen earlier.  It is apparent that for these data,
the values of $R23$ are fairly insensitive to log(O/H) as we saw before,
owing to the location of the inflection and spread in $U$.

To estimate a theoretical calibration for $S23$, we take
the mean of the three models at each metallicity, up to $0.5\Zsol$.
A resulting power-law fit is shown by the lighter, straight, solid line in
Figure~\ref{DSfig}$a$.  For $S23$, we obtain:
\begin{equation}\label{calibS23}
\log ({\rm S/H}) = -5.43 + 1.33\ \log S23 \quad .
\end{equation}
The light dash-dot and dashed lines in Figure~\ref{DSfig}$a$ show the
DPM and Christensen {\etal}(1997) calibrations, respectively.  DPM
calibrated a relation for log(O/H) vs. $S23$, so we used the
solar S/O ratio to convert their relation to a calibration of log(S/H).
It is apparent that all three calibrations are similar.  The DPM
relation shows the best correspondence to the data, as is expected
since it is fitted to the largest dataset.
It is especially encouraging that our theoretical relation is
intermediate between the two empirical ones, confirming that the
theoretical calibration is fully consistent with the available data.  
However, in using any of these $S23$ calibrations, it is
important to bear in mind that the models predict a double-valued
relation around $\log S23 \gtrsim 0.0$.

We use the same procedure to fit a theoretical calibration for $S234$
and obtain: 
\begin{equation}\label{calibS234}
\log \rm(S/H) = -5.58 + 1.27\ \log S234 \quad .
\end{equation}
As in the case for $S23$, the data are in excellent agreement with this rough
theoretical fit, shown by the light, solid line in Figure~\ref{DSfig}$b$.

We again emphasize that, although the data at $Z> 0.5\Zsol$ are
consistent with the calibrations for both $S23$ and $S234$,
they strongly diverge from the models in that regime, and that a
power-law approximation is necessarily crude near these values.
{\it We therefore consider the calibrations reliable only for} $Z \lesssim
0.5\Zsol$, and extreme caution should be exercised in extrapolating at
higher metallicity.  It is also essential to note that the double-valued
nature of all of these abundance parameters still remains an issue.

In Figure~\ref{DSfig}$d$, we show the measured log(O/H) vs. $\log
S234$ for the Dennefeld \& Stasi\'nska sample, where log(O/H)
are again derived from direct measurements of $T$\oiii.  We see that
the scatter is much larger than for log(S/H) vs. $\log S234$
(Figure~\ref{DSfig}$b$).  Although Garnett (1989), among others,
suggests that there is no systematic variation in S/O with O/H,
Figure~\ref{DSfig} shows that there is still significant 
variation in the S/O ratio among the different objects.
Therefore, while $S234$ appears reasonably reliable for estimating the S
abundance, it appears to be significantly less reliable for inferring
the O abundance, and caution should be exercised accordingly.

\section{Conclusion}

We have carried out a detailed investigation of elemental abundance
derivations using four \hii\ regions in the LMC.  We use tailored
photoioinzation models to examine standard abundance analyses based on
measured values of \te.  Our data (Paper~I) are derived from both
spatially-resolved observations extracted from stationary long slit
positions, and scanned, spatially-integrated slit observations.  We
also examine the bright-line abundance diagnostics for O and S, 
in light of the direct abundance determinations and photoionization
models.   

Our abundance determinations are based on measurements of $T$\oiii,
which we take to represent \thi, and we assume a two-zone temperature
structure for the nebulae, represented by \thi\ and \tlo.  We use
standard ionic abundance relations to then determine the total
elemental abundances for He, N, O, 
Ne, S, and Ar, with respect to H.  Comparison with tailored
{\sc Mappings} photoionization models highlights the importance of
choosing a relation between \thi\ and \tlo\ that adequately represents
the nebular temperature structure.  Failure to do so can result in
metallicity estimates that are discrepant by at least 0.2 dex from
values indicated by tailored photoionization models.

Abundance measurements for the stationary slit positions show high
spatial uniformity, with no evidence of variations or gradients to
within 0.1 -- 0.15 dex.  Thus it is unlikely that there are systematic
biases resulting from the strong ionization gradients seen in these
objects.  The adopted two-zone \te\ structure therefore appears to be
highly reliable for estimating ionic abundance estimates even through 
``pencil-beam'' apertures that sample only a small nebular area, at
least for our fairly isothermal \hii\ regions.

No areas of local enrichment were detected in DEM L199, in spite of
the presence of two WN3 stars and one WC4 star.
The stellar products may be hidden in hot,
coronal gas within the central superbubble, or the stars may not have
produced enough enriched material to be readily detectable.  The
results show that self-enrichment by WR stars is likely to be a
complex phenomenon, empirically.  DEM L243 and DEM L301, both showing
evidence of recent SNR activity, also do not show local enrichments,
although with poorer constraints.

Abundance measurements from the scanned, spatially integrated
apertures are consistent with those obtained from the spatially
resolved observations.  Our results are $\sim 0.2$ dex lower than
average LMC \hii\ region measurements (Dufour 1984; Garnett 1999),
probably resulting in part from different descriptions for the \te\
structure.  The spatially-integrated measurements are also consistent
with there being no variation between the four \hii\ regions, although,
interestingly, they are also consistent with the marginal abundance
gradient suggested by Pagel {\etal}(1978).  While the presence of the
SNR in DEM L243 did not affect the resulting abundances from the
spatially integrated observation, the derived abundances for DEM L301
are on the low end of the distribution, hinting at spurious effects
caused by the shock activity in that superbubble.

We computed the $R23$ O abundance parameters (Pagel {\etal}1979) for
the spatially integrated data, and compared these with model tracks
constructed with {\sc Mappings}.  The models assume the Costar C2
(Schaerer \& de Koter 1997) stellar atmosphere corresponding to an O6
-- O7 spectral type, and an inner nebular radius of 0.4 $R_{\rm S}$.
As has historically been the case, the models do not agree well with
the locus of observations in the literature, although our LMC data do
agree well, coincidentally, with both.  

Similarly, we examined the $S23$ abundance parameter for S (e.g., Christensen
{\etal}1997; D\'\i az \& P\'erez-Montero 2000).  Our models reveal
that, contrary to previous suggestions, $S23$ is {\it more} sensitive
to the ionization parameter than is $R23$.  \ion{S}{4} is produced by the
same radiation that ionizes \ion{O}{3}, and is a significant ion
of S in many \hii\ regions, but it is not sampled by $S23$.  Its
omission therefore causes $S23$ to be much more sensitive to $U$ than
$R23$.  The spatially resolved observations confirm this by showing,
in agreement with model predictions, lower values of $S23$ in the
central nebular regions where \ion{S}{4} is important.  As shown in
Paper~I, this spatial variation is not predicted or observed in
$R23$.  

Our models also suggest that the maximum in $S23$ occurs at
only $\sim 0.2$ dex higher in $Z$ than in $R23$.  Nevertheless, this
appears to significantly alleviate the effect of the double-valued
structure of log(S/H) vs $S23$ when inferring abundances, as
shown by D\'\i az \& P\'erez-Montero (2000).  It is highly encouraging
that the data, both from our sample and from the literature, are in
excellent agreement with the models, in contrast to the behavior of
$R23$.  We offer a theoretical calibration of $S23$
(equation~\ref{calibS23}) which appears to
be fully compatible with the data in the literature thus far.
However, we caution that the locus of the available data
may well be deceptive in suggesting that a power-law
relation can be used at $Z\gtrsim 0.5\Zsol$.

To overcome the limitations of $S23$ in $U$-sensitivity and spatial
variation, we introduce a similar S abundance parameter, $S234$.  This
is the same as $S23$ with the added emission of
\siv\lam10.5$\mu$.  Although this mid-IR line is not readily
observable with most conventional ground-based spectrographs, it is
straightforward to estimate its intensity from the simple
correspondence between \siv/\siii\ and \oiii/\oii\
(equation~\ref{eqS4O3}).  Our models show that $S234$ is 
less dependent on $U$ than is even $R23$.  $S234$ for our objects and for
the larger sample of Dennefeld \& Stasi\'nska (1983) are in excellent
agreement with the models.  Likewise, the spatial variations for both
models and observations are dramatically reduced for
$S234$ in contrast to $S23$.  We provide a theoretical calibration for
log(S/H) vs $S234$ at $Z\lesssim 0.5\Zsol$ (equation~\ref{calibS234}).  

Finally, we reiterate some caveats for the use of $R23$, $S23$, and
$S234$.  We find that the presence of shock excitation increases the
value of these parameters; for our objects, the effect is about 0.1
dex in magnitude.  Secondly, significant variations in the S/O ratio dictate
caution in inferring O abundances using $S234$ and $S23$
(Figure~\ref{DSfig}).  It is also important to bear in mind the
double-valued structure for all three of these parameters.
Lastly, we emphasize the deviation between the
data and models above $0.5\Zsol$, and we therefore consider
the calibrations presented thus far for $S23$ and $S234$ to be
reliable only for $Z \lesssim 0.5\Zsol$.  Further empirical
investigation is needed to understand the behavior of these parameters
at higher metallicity.  Bearing in mind these caveats, the excellent
correspondence between the modeled $S234$, $S23$, and 
the available data, together with the more highly monotonic
behavior of these parameters, promises greater effectiveness as
metallicity indicators than $R23$.  With improving access to
the \siv\lam10.5$\mu$ line, it should be possible to confirm
the behavior of $S234$ directly.

\acknowledgments

It is a pleasure to acknowledge discussions with Mike Dopita, Annette
Ferguson, Don Garnett, Dick Shaw, Evan Skillman, Elena
Terlevich, and Bob Williams.  We are also grateful to Mike Dopita for
access to the {\sc Mappings~II} photoionization code and to Angelez
D\'\i az for access to her work in advance of publication.  Finally, we 
are pleased to acknowledge the referee, Bernard Pagel.


\begin{deluxetable}{clllllll}
\footnotesize
\tablecaption{\hii\ region sample \label{sample}}
\tablewidth{0 pt}
\tablehead{
\colhead{DEM} & \colhead{Henize}   & \colhead{LH} & \colhead{Sp. Type}
& \colhead{$\log (L_{\ha}/\ergs)$ \tablenotemark{a}} & \colhead{Inner radius}
& \colhead{Shocks} & \colhead{Sp. Type Reference}
} 
\startdata
DEM L199 & N144 & LH 58    &  WN3 & 38.6 & 
	0.5 $R_{\rm S}$\tablenotemark{b} & no & Breysacher (1981) \nl
DEM L243 & N63 A & LH 83   &  O7  & 37.7 & 
	0.1 $R_{\rm S}$ & yes & Oey (1996a) \nl
DEM L301 & N70 & LH 114    &  O3  & 37.7\tablenotemark{c} & 
	0.95 $R_{\rm S}$ & yes & Oey (1996a) \nl
DEM L323 & N180 B & LH 117 &  O3--4 & 38.4\tablenotemark{d} & 
	0.4 $R_{\rm S}$ & no & Massey {\etal}(1989) \nl
\enddata
\tablenotetext{a}{From Oey \& Kennicutt (1997), unless otherwise indicated.}
\tablenotetext{b}{DEM L199 has complex morphology (see Paper~I).}
\tablenotetext{c}{DEM L301 is significantly density-bounded (Paper~I).}
\tablenotetext{d}{Kennicutt (1997), private communication.}
\end{deluxetable}

\begin{deluxetable}{lcccccccccccc}
\scriptsize
\tablecaption{Logarithmic abundances of spatially resolved 
	observations\tablenotemark{a}	\label{stadat}}
\tablewidth{0 pt}
\tablehead{
\colhead{Observation} & \colhead{[He/H]\tablenotemark{b}} 
& \colhead{err} & 
\colhead{[N/H]} & \colhead{err} & 
\colhead{[O/H]} & \colhead{err} & 
\colhead{[Ne/H]\tablenotemark{c}} & \colhead{err} & 
\colhead{[S/H]} & \colhead{err} & 
\colhead{[Ar/H]} & \colhead{err}
} 
\startdata
D199.205-ap10    & --1.059  & 0.026 	 & --4.892  & 0.126   & --3.647  & 0.216   & --4.390  & 0.256   & --5.282  & 0.116   & --5.993  & 0.133 \\
D199.205-ap11    & --1.091  & 0.021 	 & --4.899  & 0.044   & --3.633  & 0.065   & --4.374  & 0.076   & --5.261  & 0.038   & --5.896  & 0.046 \\
D199.205-ap13    & --1.039  & 0.015 	 & --4.890  & 0.040   & --3.621  & 0.056   & --4.368  & 0.064   & --5.280  & 0.032   & --5.889  & 0.040 \\
D199.205-ap14    & --1.066  & 0.020 	 & --4.977  & 0.076   & --3.571  & 0.130   & --4.409  & 0.158   & --5.172  & 0.063   & --5.583  & 0.080 \\
D199.205-ap15    & --1.057  & 0.026 	 & --4.927  & 0.118   & --3.681  & 0.204   & --4.483  & 0.241   & --5.378  & 0.105   & --5.886  & 0.125 \\
D199.205-ap16    & --1.016  & 0.027 	 & --4.946  & 0.175   & --3.421  & 0.313   & --4.080  & 0.372   & --5.315  & 0.157   & --5.790  & 0.186 \\
D199.205-ap17    & --1.081  & 0.029 	 & --4.947  & 0.104   & --3.689  & 0.182   & --4.537  & 0.225   & --5.315  & 0.096   & --5.924  & 0.113 \\
D199.205-ap18    & --1.001  & 0.069 	 & --4.545  & 0.327   & --3.018  & 0.587   & --3.635  & 0.706   & --4.958  & 0.308   & --5.577  & 0.358 \\
D199.205N120-ap10    & --1.065  & 0.029 	 & --4.928  & 0.178   & --3.669  & 0.302   & --4.316  & 0.359   & --5.316  & 0.161   & --6.010  & 0.186 \\
D199.205N120-ap11    & --1.065  & 0.027 	 & --4.932  & 0.137   & --3.627  & 0.237   & --4.337  & 0.283   & --5.337  & 0.125   & --5.960  & 0.147 \\
D199.205N120-ap12    & --1.083  & 0.015 	 & --4.999  & 0.043   & --3.771  & 0.063   & --4.553  & 0.074   & --5.391  & 0.037   & --6.104  & 0.047 \\
D199.205N120-ap13    & --1.088  & 0.018 	 & --4.950  & 0.046   & --3.695  & 0.072   & --4.423  & 0.085   & --5.337  & 0.041   & --5.992  & 0.049 \\
D199.205N120-ap14    & --1.011  & 0.015 	 & --4.898  & 0.047   & --3.552  & 0.073   & --4.317  & 0.085   & --5.268  & 0.040   & --5.814  & 0.049 \\
D199.205N120-ap15    & --1.033  & 0.019 	 & --4.942  & 0.077   & --3.573  & 0.130   & --4.344  & 0.154   & --5.316  & 0.068   & --5.867  & 0.081 \\
D199.205N120-ap16    & --1.031  & 0.019 	 & --4.869  & 0.089   & --3.569  & 0.154   & --4.347  & 0.184   & --5.259  & 0.081   & --5.815  & 0.097 \\
D199.205N120-ap17    & --1.033  & 0.027 	 & --5.205  & 0.161   & --3.522  & 0.296   & --4.271  & 0.334   & --5.264  & 0.118   & --5.674  & 0.148 \\
D199.205N120-ap18    & --1.046  & 0.027 	 & --4.905  & 0.141   & --3.382  & 0.252   & --4.071  & 0.301   & --5.258  & 0.128   & --5.755  & 0.151 \\
D199.205N120-ap19    & --1.014  & 0.024 	 & --4.909  & 0.131   & --3.578  & 0.230   & --4.304  & 0.275   & --5.218  & 0.119   & --5.763  & 0.143 \\
D243.2S-ap6   & --1.068  & 0.026   	 & --5.412  & 0.110   & --4.177  & 0.175   & --5.306  & 0.242   & --5.710  & 0.095   & --6.360  & 0.115\\ 
D243.2S-ap9   & --1.176  & 0.025   	 & --5.279  & 0.151   & --4.094  & 0.257   & --5.064  & 0.303   & --5.683  & 0.136   & --6.393  & 0.158 \\
D243.2S-ap14    & --1.313  & 0.031 	 & --5.252  & 0.108   & --4.055  & 0.173   & --4.940  & 0.206   & --5.643  & 0.098   & --6.570  & 0.119 \\
D243.5S-ap3   & --1.559  & 0.069   	 & --5.408  & 0.350   & --4.337  & 0.563   & --5.442  & 0.775   & --5.801  & 0.317   & --7.030  & 0.338 \\
D243.5S-ap6   & --1.296  & 0.023   	 & --5.079  & 0.177   & --3.703  & 0.305   & --4.675  & 0.363   & --5.414  & 0.159   & --6.200  & 0.182 \\
D243.5S-ap5   & --1.101  & 0.019   	 & --5.074  & 0.064   & --3.587  & 0.107   & --4.457  & 0.130   & --5.375  & 0.056   & --5.931  & 0.066 \\
D243.5S-ap7   & --1.156  & 0.034   	 & --4.843  & 0.290   & --3.315  & 0.507   & --4.215  & 0.607   & --5.246  & 0.262   & --5.885  & 0.297 \\
D243.5S-ap8   & --1.161  & 0.039   	 & --5.085  & 0.263   & --3.634  & 0.452   & --4.654  & 0.556   & --5.406  & 0.235   & --6.073  & 0.276 \\
D243.30S-ap9   & --1.293  & 0.125   	 & --5.412  & 0.319   & --4.397  & 0.520   & --5.678  & 0.720   & --5.803  & 0.298   & --6.617  & 0.323 \\
D243.30S-ap10    & --1.146  & 0.047 	 & --5.219  & 0.189   & --3.944  & 0.317   & --4.951  & 0.400   & --5.526  & 0.169   & --6.242  & 0.198 \\
D243.30S-ap11    & --1.164  & 0.036 	 & --4.763  & 0.298   & --3.147  & 0.522   & --4.079  & 0.641   & --5.191  & 0.270   & --5.769  & 0.312 \\
D243.30S-ap14    & --1.257  & 0.077 	 & --5.413  & 0.240   & --4.332  & 0.397   & --5.398  & 0.524   & --5.786  & 0.222   & --6.594  & 0.254 \\
D243.30S-ap15    & --1.108  & 0.098 	 & --4.587  & 0.320   & --2.909  & 0.549   & --3.563  & 0.675   & --4.980  & 0.292   & --5.608  & 0.340 \\
D243fix.2S-ap6   & --1.083  & 0.025   	 & --5.145  & 0.065   & --3.745  & 0.100   & --4.796  & 0.150   & --5.476  & 0.054   & --6.072  & 0.068 \\
D243fix.2S-ap9   & --1.185  & 0.024   	 & --5.083  & 0.055   & --3.778  & 0.089   & --4.687  & 0.116   & --5.508  & 0.048   & --6.180  & 0.058 \\
D243fix.2S-ap14    & --1.332  & 0.030 	 & --4.991  & 0.056   & --3.633  & 0.091   & --4.441  & 0.114   & --5.405  & 0.051   & --6.289  & 0.071 \\
D243fix.5S-ap3   & --1.575  & 0.079   	 & --5.085  & 0.057   & --3.813  & 0.091   & --4.824  & 0.411   & --5.513  & 0.048   & --6.693  & 0.095 \\
D243fix.5S-ap6   & --1.295  & 0.022   	 & --5.095  & 0.054   & --3.731  & 0.089   & --4.708  & 0.109   & --5.428  & 0.049   & --6.216  & 0.058 \\
D243fix.5S-ap5   & --1.096  & 0.019   	 & --5.167  & 0.055   & --3.755  & 0.087   & --4.657  & 0.109   & --5.459  & 0.049   & --6.031  & 0.058 \\
D243fix.5S-ap7   & --1.143  & 0.031   	 & --5.096  & 0.056   & --3.757  & 0.090   & --4.737  & 0.135   & --5.473  & 0.053   & --6.144  & 0.066 \\
D243fix.5S-ap8   & --1.158  & 0.037   	 & --5.130  & 0.059   & --3.713  & 0.089   & --4.747  & 0.160   & --5.447  & 0.052   & --6.120  & 0.075 \\
D243fix.30S-ap9   & --1.305  & 0.132   	 & --5.010  & 0.058   & --3.748  & 0.092   & --4.919  & 0.432   & --5.446  & 0.055   & --6.196  & 0.090 \\
D243fix.30S-ap10    & --1.150  & 0.047 	 & --5.078  & 0.056   & --3.715  & 0.089   & --4.678  & 0.176   & --5.400  & 0.050   & --6.089  & 0.063 \\
D243fix.30S-ap11    & --1.151  & 0.033 	 & --5.086  & 0.054   & --3.722  & 0.087   & --4.760  & 0.192   & --5.484  & 0.050   & --6.107  & 0.060 \\
D243fix.30S-ap14    & --1.283  & 0.076 	 & --5.089  & 0.059   & --3.810  & 0.092   & --4.783  & 0.278   & --5.493  & 0.059   & --6.247  & 0.075 \\
D243fix.30S-ap15    & --1.084  & 0.097 	 & --5.063  & 0.065   & --3.759  & 0.093   & --4.573  & 0.236   & --5.421  & 0.059   & --6.105  & 0.092 \\
D301.SW6-ap7   & --1.061  & 0.019   	 & --5.193  & 0.059   & --3.984  & 0.092   & --4.695  & 0.111   & --5.487  & 0.052   & --6.323  & 0.063 \\
D301.SW6-ap8   & --1.031  & 0.020   	 & --5.116  & 0.062   & --3.874  & 0.097   & --4.672  & 0.118   & --5.445  & 0.055   & --6.205  & 0.067 \\
D301.SW6-ap9   & --1.065  & 0.020   	 & --4.975  & 0.199   & --3.667  & 0.336   & --4.377  & 0.397   & --5.279  & 0.178   & --6.052  & 0.199 \\
D301.SW6-ap10    & --0.920  & 0.025 	 & --5.071  & 0.064   & --3.820  & 0.101   & --4.475  & 0.120   & --5.411  & 0.057   & --6.144  & 0.070 \\
D301.SW6-ap11    & --1.036  & 0.020 	 & --5.163  & 0.062   & --3.912  & 0.098   & --4.743  & 0.125   & --5.439  & 0.055   & --6.228  & 0.067 \\
D301.SW1-ap6   & --1.135  & 0.032   	 & --5.008  & 0.309   & --3.710  & 0.514   & --4.190  & 0.607   & --5.362  & 0.277   & --6.254  & 0.305 \\
D301.SW1-ap5   & --1.088  & 0.018   	 & --4.980  & 0.195   & --3.518  & 0.334   & --4.437  & 0.395   & --5.297  & 0.169   & --5.984  & 0.193 \\
%
%
D323.C2-ap7   & --1.194  & 0.022   	 & --5.127  & 0.105   & --3.597  & 0.180   & --4.519  & 0.213   & --5.327  & 0.090   & --5.958  & 0.109 \\
D323.C2-ap8   & --1.089  & 0.017   	 & --5.168  & 0.070   & --3.710  & 0.114   & --4.625  & 0.135   & --5.421  & 0.060   & --6.040  & 0.074 \\
D323.C2-ap9   & --1.072  & 0.016   	 & --5.084  & 0.048   & --3.594  & 0.078   & --4.437  & 0.091   & --5.349  & 0.042   & --5.942  & 0.051 \\
D323.C2-ap10    & --1.190  & 0.021 	 & --5.113  & 0.042   & --3.627  & 0.060   & --4.483  & 0.069   & --5.323  & 0.035   & --5.908  & 0.042 \\
D323.C2-ap11    & --1.049  & 0.016 	 & --5.149  & 0.058   & --3.639  & 0.089   & --4.488  & 0.102   & --5.316  & 0.046   & --5.843  & 0.056 \\
D323.C2-ap12    & --1.057  & 0.016 	 & --5.127  & 0.053   & --3.631  & 0.082   & --4.511  & 0.096   & --5.294  & 0.043   & --5.883  & 0.051 \\
D323.C2-ap13    & --1.066  & 0.027 	 & --5.268  & 0.083   & --3.852  & 0.130   & --4.782  & 0.159   & --5.419  & 0.072   & --6.074  & 0.088 \\
D323.C1-ap13    & --1.080  & 0.051 	 & --4.995  & 0.302   & --3.461  & 0.526   & --4.341  & 0.639   & --5.265  & 0.262   & --5.885  & 0.313 \\
D323.C1-ap6   & --1.169  & 0.019   	 & --4.919  & 0.127   & --3.361  & 0.219   & --4.255  & 0.259   & --5.207  & 0.108   & --5.912  & 0.127 \\
D323.C1-ap7   & --1.175  & 0.022   	 & --5.238  & 0.053   & --3.660  & 0.088   & --4.575  & 0.104   & --5.345  & 0.046   & --5.891  & 0.054 \\
D323.C1-ap8   & --1.050  & 0.016   	 & --5.178  & 0.052   & --3.635  & 0.082   & --4.513  & 0.095   & --5.306  & 0.043   & --5.835  & 0.051 \\
D323.C1-ap9   & --1.077  & 0.016   	 & --5.165  & 0.033   & --3.602  & 0.044   & --4.454  & 0.050   & --5.361  & 0.028   & --5.920  & 0.032 \\
D323.C1-ap10    & --1.085  & 0.017 	 & --5.112  & 0.067   & --3.618  & 0.112   & --4.525  & 0.132   & --5.359  & 0.058   & --5.993  & 0.070 \\
D323.C1-ap11    & --1.059  & 0.016 	 & --5.133  & 0.040   & --3.678  & 0.063   & --4.763  & 0.080   & --5.316  & 0.036   & --5.995  & 0.042 \\
D323.C1-ap12    & --1.113  & 0.029 	 & --5.149  & 0.253   & --3.713  & 0.425   & --4.530  & 0.513   & --5.391  & 0.217   & --6.159  & 0.252 \\
\enddata
\tablenotetext{a}{D243fix observations have values computed for \thi\
	fixed at 9700 K.}
\tablenotetext{b}{[He/H] are lower limits for DEM L243.}
\tablenotetext{c}{[Ne/H] is subject to systematic uncertainties
	(see text).}
\end{deluxetable}

\begin{deluxetable}{lcccccccccccccc}
\scriptsize
\tablecaption{Mean logarithmic abundances from spatially resolved observations 
	\label{meandat}}
\tablewidth{0 pt}
\tablehead{
\colhead{Object} & {$T$[O III](K)} & \colhead{$\sigma$} & 
\colhead{[He/H]} & \colhead{$\sigma$} & 
\colhead{[N/H]} & \colhead{$\sigma$} & 
\colhead{[O/H]} & \colhead{$\sigma$} & 
\colhead{[Ne/H]\tablenotemark{a}} & \colhead{$\sigma$} & 
\colhead{[S/H]} & \colhead{$\sigma$} & 
\colhead{[Ar/H]} & \colhead{$\sigma$}
} 
\startdata
DEM L199 & 9620 & 140 & --1.05 & 0.005 & --4.93 & 0.017 & --3.64 & 0.026 & --4.40 & 0.030 & --5.30 & 0.014 & --5.90 & 0.018 \\
DEM L243 & 9620 & 360 & --1.18\tablenotemark{b} & 0.009 & --5.17 & 0.040 & --3.83 & 0.068 & --4.75 & 0.084 & --5.50 & 0.036 & --6.18 & 0.043 \\
DEM L243\tablenotemark{c} & 9700 & \nodata 
	& --1.18\tablenotemark{b} & 0.009 & --5.09 & 0.016 & --3.74 & 0.025 & --4.67 & 0.042 & --5.46 & 0.014 & --6.17 & 0.019 \\
DEM L301 & 11900 & 420 & --1.05 & 0.008 & --5.13 & 0.030 & --3.89 & 0.047 & --4.63 & 0.058 & --5.44 & 0.027 & --6.22 & 0.032 \\
DEM L323 & 9480 & 110 & --1.09 & 0.005 & --5.15 & 0.015 & --3.63 & 0.022 & --4.52 & 0.026 & --5.33 & 0.012 & --5.93 & 0.015 \\
\enddata
\tablenotetext{a}{[Ne/H] is subject to systematic uncertainties
	(see text).}
\tablenotetext{b}{[He/H] is a lower limit in this object.}
\tablenotetext{c}{\thi\ fixed at 9700 K for all apertures.}
\end{deluxetable}

\begin{deluxetable}{lcccccccccccccc}
\scriptsize
\tablecaption{Logarithmic abundances from spatially integrated observations 
	\label{scandat}}
\tablewidth{0 pt}
\tablehead{
\colhead{Observation} & $T$[O III](K) & \colhead{err} & 
\colhead{[He/H]} & \colhead{err} & 
\colhead{[N/H]} & \colhead{err} & 
\colhead{[O/H]} & \colhead{err} & 
\colhead{[Ne/H]\tablenotemark{a}} & \colhead{err} & 
\colhead{[S/H]} & \colhead{err} & 
\colhead{[Ar/H]} & \colhead{err}
} 
\startdata
D199.496W240 & 8100 & 1100 & --1.13 & 0.03 & --4.77 & 0.17 & --3.31 & 0.31 & --3.97 & 0.37 & --5.19 & 0.16 & --5.73 & 0.18 \\
D243.2(total) & 11900 & 1000 & --1.18 & 0.02 & --5.24 & 0.08 & --4.01 & 0.12 & --4.90 & 0.14 & --5.54 & 0.07 & --6.36 & 0.08 \\
D243.2(no SNR) & 11400 & 1400 & --1.17 & 0.03 & --5.26 & 0.12 & --3.97 & 0.19 & --4.90 & 0.23 & --5.58 & 0.10 & --6.26 & 0.12 \\
D301.SW6 & 13000 & 900 & --1.04 & 0.02 & --5.19 & 0.06 & --3.94 & 0.09 & --4.66 & 0.10 & --5.50 & 0.05 & --6.27 & 0.06 \\
D323.140(total) & 9600 & 700 & --1.09 & 0.02 & --5.10 & 0.08 & --3.64 & 0.14 & --4.54 & 0.17 & --5.35 & 0.07 & --5.97 & 0.09 \\
D323.140N30N & 9700 & 1400 & --1.11 & 0.03 & --5.13 & 0.18 & --3.66 & 0.30 & --4.57 & 0.36 & --5.35 & 0.15 & --6.04 & 0.18 \\
D323.140N30S & 8800 & 400 & --1.21 & 0.02 & --5.02 & 0.06 & --3.48 & 0.10 & --4.30 & 0.12 & --5.27 & 0.05 & --5.85 & 0.06 \\
\hline\\
Dufour (1984)\tablenotemark{b}& & & --1.07 & 0.02 & --5.03 & 0.10 & --3.57 & 0.08 & --4.36 & 0.10 & --5.15 & 0.11 & --5.80 & 0.06 \\
\enddata
\tablenotetext{a}{[Ne/H] is subject to systematic uncertainties
	(see text).}
\tablenotetext{b}{Mean LMC \hii\ region abundances as compiled by
	Dufour (1984).}
\end{deluxetable}

\begin{deluxetable}{lcccccc}
\footnotesize
\tablecaption{Abundance parameters for spatially integrated observations 
	\label{tababdpar}}
\tablewidth{0 pt}
\tablehead{
\colhead{Observation} & \colhead{$R23$\tablenotemark{a}} & \colhead{err} & 
\colhead{$S23$\tablenotemark{a}} & \colhead{err} & 
\colhead{$S234$\tablenotemark{b}} & \colhead{err}
} 
\startdata
D199.496W240 & 6.9 & 0.4 & 1.11 & 0.08 & 1.38 & 0.38 \\
D243.2 (total) & 4.7 & 0.5 & 1.31 & 0.08 & 1.35 & 0.36 \\
D243.2 (no SNR) & 4.7 & 0.5 & 1.04 & 0.07 & 1.09 & 0.29 \\
D301.SW6 & 7.1 & 0.9 & 1.83 & 0.11 & 1.85 & 0.49 \\
D323.140 & 6.1 & 0.5 & 1.19 & 0.08 & 1.31 & 0.34 \\
D323.140N30N & 5.8 & 0.6 & 1.27 & 0.09 & 1.34 & 0.35 \\
D323.140N30S & 6.2 & 0.5 & 1.15 & 0.08 & 1.29 & 0.34 \\
\enddata
\tablenotetext{a}{From Paper~I.}
\tablenotetext{b}{Computed by estimating \siv\ with equation~\ref{eqS4O3}.}
\end{deluxetable}

%
%

\clearpage

\textheight=8.4in
\topmargin=0in
\headheight=.15in



%
%

\end{document}